\documentclass[twocolumn,pre,aps,showpacs,amsmath,amssymb]{revtex4-1}
\usepackage{graphicx}
\usepackage{bm}
\usepackage{soul}
\usepackage{color}

\begin{document}

\title{Electrically-controlled self-similar evolution of viscous fingering patterns in radial Hele-Shaw flows}
\author{Pedro H. A. Anjos$^{1}$}
\email[]{pamorimanjos@iit.edu (corresponding author)}
\author{Meng Zhao$^{2}$}
\email[]{mzhao9@hust.edu.cn}
\author{John Lowengrub$^{3}$}
\email[]{lowengrb@math.uci.edu}
\author{Shuwang Li$^{1}$}
\email[]{sli@math.iit.edu\\ These authors contributed equally: Pedro H. A. Anjos, Meng Zhao.}
\affiliation{$^{1}$Department of Applied Mathematics, Illinois Institute of Technology,
	Chicago, Illinois 60616, USA \\ 
	$^2$ Center for Mathematical Sciences, Huazhong University of Science and Technology, Wuhan 430074, China \\
$^3$Department of Mathematics, University of California at Irvine, Irvine, California 92697, USA}

%\date{\today}

\begin{abstract}

Time-dependent injection strategies are commonly employed to control the number of viscous fingers emerging at the interface separating two fluids during radial displacement in Hele-Shaw flows. Here we demonstrate theoretically that such a usual controlling method is significantly improved by taking advantage of an electro-osmotic flow generated by applying an external electric field. More specifically, under the coupled action of time-varying electric currents and injection rates, we design a strategy capable of controlling not only the number of fingers emerging at the interface but also when (and if) the self-similar evolution occurs. In addition, the level of instability of the $n$-fold fingered patterns can also be tuned. This improved control over the interfacial features cannot be realized by the sole consideration of a time-varying injection rate. Perturbative second-order mode-coupling analysis and boundary integral simulations confirm that the validity and effectiveness of the controlling protocol go beyond the linear regime.

\end{abstract}
%\pacs{47.15.gp, 47.54.-r, 47.20.Ma, 47.15.km}
\maketitle

\section*{Introduction}
\label{intro}

Viscous fingering (or Saffman-Taylor instability)~\cite{Saf} is perhaps the most well-known and studied phenomenon among a family of phenomena that exhibit interfacial instabilities. This hydrodynamic instability arises when a fluid displaces another of higher viscosity in the narrow gap separating two flat, parallel glass plates of an experimental device known as Hele-Shaw cell. In its conventional radial setup~\cite{Lp}, the less viscous fluid is injected at a constant injection rate at the center of the cell, working against viscous and surface tension forces to drive the more viscous fluid radially outwards. The interplay of these forces is responsible for the formation of highly intricate interfacial patterns~\cite{Rev1,Rev2,Rev3}, characterized by their typical branching due to tip-splitting and finger competition behaviors.

Despite the rich dynamical behaviors and eye-catching morphologies, the emergence of intricate patterned structures can be detrimental for some technological and industrial applications~\cite{Gorell,Stokes,Anke,Nase,pedroadh}, due to their unpredictable, disordered growth. For instance, processes related to oil recovery~\cite{Gorell,Stokes} by water flooding method are very inefficient if viscous fingering develops at the interface separating the water and oil phases. On the other hand, it is well-known that the emergence of these instabilities enhances fluids mixing~\cite{Juanes} and is therefore desirable in that case. These facts have stimulated research efforts to develop a fundamental understanding of the interfacial dynamics of these systems and to find ways to control such hydrodynamic instabilities. So, methods aimed toward controlled suppression or enhancement of fingering instabilities or strategies capable of prescribing an ordered growth of viscous fingers are of technological and scientific importance.

Several controlling strategies have been developed in the past years, exploiting, for instance, the manipulation of the Hele-Shaw cell geometry~\cite{Stone1,Bon,pedroGEOM,Liam,novoo1,novoo2}, usage of elastic-walled cells~\cite{Draga12}, and employment of specific time-dependent injection fluxes~\cite{ShuwangPRL,mama1,mama2,mama4,irio,woods,novoo3,novoo4,novoo5,pedrowet} and gap widths~\cite{pedrowet,Juel,Zhao18,Stone,mama3}. More recently, researchers~\cite{Bazant1,Bazant2} have achieved interfacial control in Hele-Shaw cells by harnessing electro-osmotic flows generated through externally applied electric fields. Electro-osmotic flow arises over electrically charged surfaces due to the interaction of an externally applied electric field with the net charge in the electric double layer. In the context of Hele-Shaw flows, this electric double layer is a thin region formed by an accumulation of ions in the liquid attracted by the charged surface of the glass plates of the Hele-Shaw. In Refs.~\cite{Bazant1,Bazant2}, the authors demonstrated via linear stability analysis, numerical simulations, and experiments that, depending on the magnitude and direction of the electric current, viscous fingering instabilities can be either enhanced or suppressed.

Here, we further explore the system examined in Refs.~\cite{Bazant1,Bazant2}, but with a different goal in mind: Instead of using electric fields to enhance or suppress interfacial instabilities, we utilize it to promote controlled evolution of an unstable interface and also to tune interesting nonlinear behaviors. More specifically, we design a controlling strategy coupling time-dependent electric currents with time-varying injection rates that can make the interface evolve with a fixed number of fingers, suppressing the formation of branched patterns commonly observed in unstable Hele-Shaw flows. This was achieved previously in Refs.~\cite{ShuwangPRL,mama1,irio,woods,pedrowet} by utilizing only a time-dependent injection rate. However, by adding the electric-osmotic contribution to the pressure-driven flow, we demonstrate that the role of fixing the number of fingers is transferred from the injection rate to the applied electric current. Therefore, while the time-dependent electric current is responsible for prescribing the number of fingers emerging at the interface for a given set of physical parameters and initial conditions, the injection rate can vary freely to tune when, and if, the flow evolves to a self-similar growth. In addition, one may also control the level of instability of the interface. In contrast to the previous method based solely on the usage of injection rate, in our present protocol, all of these interfacial features are controlled without altering physical parameters, initial conditions, and the number of fingers at the interface. Thus, here we present a superior method of control than previously reported strategies.

\begin{figure}[t]
	\centering
	\includegraphics[scale=0.42]{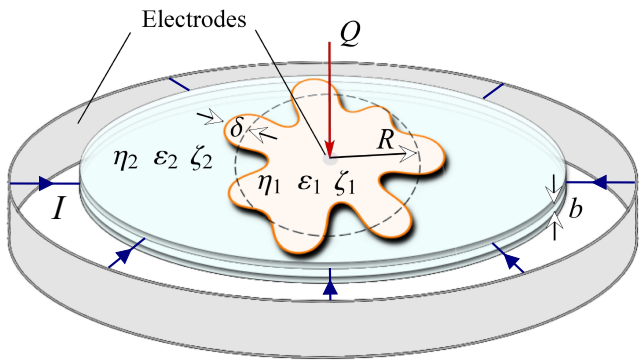} 
	\caption{{\bf Representative sketch of the electrohydrodynamic Hele-Shaw flow.} Schematic representation of the radial displacement of an outer fluid having viscosity $\eta_{2}$, permittivity $\varepsilon_{2}$, and zeta (surface) potential $\zeta_{2}$, by the injection of an inner fluid with $\eta_{1}$, $\varepsilon_{1}$, and $\zeta_{1}$. The inner fluid is injected at a volumetric flow rate $Q=Q(t)$, and the flow takes place between the plates of a Hele-Shaw cell of gap thickness $b$. The system is subjected to an electric current $I=I(t)$ produced by electrodes located at the center and at the outer edge of the Hele-Shaw cell. The time-dependent unperturbed interface radius is represented by $R=R(t)$, and small interfacial perturbations are denoted by $\delta=\delta(\theta, t)$, where $\theta$ is the azimuthal angle.}
	\label{geom}
\end{figure}

\section*{Results}
\label{tb}

\subsection*{Theoretical background}
\label{gov}

A schematic representation of the flow arrangement is illustrated in Fig.~\ref{geom}. Consider a radial Hele-Shaw cell of gap thickness $b$ containing two immiscible, incompressible, Newtonian viscous fluids. Denote the viscosities, permittivities, and zeta (surface) potentials of the inner and outer fluids as $\eta_{1}$, $\varepsilon_{1}$, $\zeta_{1}$, and $\eta_{2}$, $\varepsilon_{2}$, $\zeta_{2}$, respectively. At the interfacial boundary where the two fluids meet, there exists a surface tension $\gamma$. The inner fluid $1$ is injected at the center of the cell at a given volumetric injection rate $Q = Q(t)$, displacing the outer fluid $2$ radially. An electric current $I=I(t)$ is applied by electrodes positioned at the center and the outer edge of the Hele-Shaw cell, and an external, in-plane electric field ${\textbf E}_{j}=-\bm{\nabla}V_{j}$ is established parallel to the flow direction, where $V_{j}$ is the electric potential in fluid $j$ ($j=1, 2$). The subscripts $1$ and $2$ refer to the inner and outer fluids, respectively.

When the fluids are brought into contact with the Hele-Shaw cell's glass plates, the glass surface becomes negatively charged due to the dissociation of ionic surface groups. Consequently, a thin, diffuse cloud of excess counter-ions in the liquids accumulate near the surface, forming the so-called electric double layer (EDL)~\cite{EDL1,EDL2,EDL3}. By applying the external electric field ${\textbf E}_{j}$, these mobile ions are subjected to a net electric force, which drives an electro-osmotic flow, in addition to the pressure-driven flow. Note that this electric-induced flow can be either in the same or in the opposite direction of hydraulic flow, depending on the direction of the applied electric field, which is related to the sign of the electric current established between the two electrodes. Positive (negative) current is defined to be in the same (opposite) direction as the flow.

By considering the contribution of electric forces, the motion of the fluids in the effectively two-dimensional Hele-Shaw cell problem is described by a modified Darcy's law for the gap-averaged velocity~\cite{EDL2,Bazant1,Bazant2,Bazant3}
\begin{equation}
\label{Darcy1}
{\textbf v}_{j} =-M_{j}\bm{\nabla}p_{j} -K_{j}\bm{\nabla}V_{j},
\end{equation}
where $M_{j}=b^{2}/12 \eta_{j}$ and $K_{j}=-\varepsilon_{j}\zeta_{j}/\eta_{j}$ are the hydraulic and electro-osmotic mobilities, respectively. This specific format for $K_{j}$ is given by the Helmholtz-Smoluchowski relation~\cite{EDL2}, which is valid for very thin EDLs. In addition, the first (second) term appearing on the right-hand side of Eq.~(\ref{Darcy1}) represents the pressure-driven (electro-osmotic) contribution to the flow.  

As the electric field acts on the EDL to promote electro-osmotic flow, pressure gradients also drive advection of ions on the EDL, generating streaming current. As a consequence, the total current density ${\textbf J}_{j}$ is the sum of streaming (pressure-induced) and Ohmic (associated to the electric field) currents 
\begin{equation}
	\label{Darcy2}
	{\textbf J}_{j} =-K_{j}\bm{\nabla}p_{j} -\sigma_{j}\bm{\nabla}V_{j},
\end{equation}
where $\sigma_{j}$ is the Ohmic conductivity of the fluid $j$.

From Eqs.~(\ref{Darcy1}) and~(\ref{Darcy2}), it can be seen that the velocity and current density fields are irrotational in the bulk, i.e., ${\bm \nabla} \times {\bf v}_{j}=0$ and ${\bm \nabla} \times {\bf J}_{j}=0$. Therefore, we can conveniently state our moving boundary problem in terms of velocity $\phi_{j}$ $({\textbf v}_{j}=-\bm{\nabla} \phi_{j})$ and current density $\psi_{j}$ $({\textbf J}_{j}=-\bm{\nabla} \psi_{j})$ potentials. In addition, by considering the incompressibility conditions ${\bm \nabla}\cdot {\bf v}_{j}=0$ and ${\bm \nabla}\cdot {\bf J}_{j}=0$, one verifies that the potentials $\phi_{j}$ and $\psi_{j}$ obey the Laplace equations ${\bm \nabla}^2 \phi_{j}=0$ and ${\bm \nabla}^2 \psi_{j}=0$, respectively.

In the context of pressure-driven, electro-osmotic Hele-Shaw flows, our physical problem is specified by four boundary conditions at the fluid-fluid interface. The first one expresses the pressure discontinuity due to the interfacial surface tension $\gamma$, and it is given by the Young-Laplace pressure boundary condition~\cite{Saf,Rev1,Rev2,Rev3}
\begin{equation}
	\label{pressure}
	p_{1}-p_{2}= \gamma \kappa,
\end{equation}
where $\kappa$ denotes the curvature of the fluid-fluid interface. Conversely, the electric potential is continuous across the interface~\cite{Bazant1,Bazant2,Bazant3}
\begin{equation}
	\label{ectro}
	V_{1}-V_{2}= 0.
\end{equation}
The other two remaining fluid-fluid conditions are the kinematic boundary conditions~\cite{Saf,Rev1,Rev2,Rev3,Bazant1,Bazant2,Bazant3}, which express the fact that the normal components of the fluids' velocities and also of the current densities are continuous across the interface. These conditions are expressed as
\begin{equation}
	\label{kinematic1} 
	{\textbf v}_{1} \cdot {\bf \hat {n}}={\textbf v}_{2} \cdot {\bf \hat {n}},
\end{equation}
and 
\begin{equation}
	\label{kinematic2} 
	{\textbf J}_{1} \cdot {\bf \hat {n}}={\textbf J}_{2} \cdot {\bf \hat {n}},
\end{equation}
where ${\bf \hat {n}}$ denotes the unit normal vector at the interface.

\subsection*{Linear growth rate}
\label{WNL}

\begin{figure*}[t]
	\centering
	\includegraphics[scale=0.62]{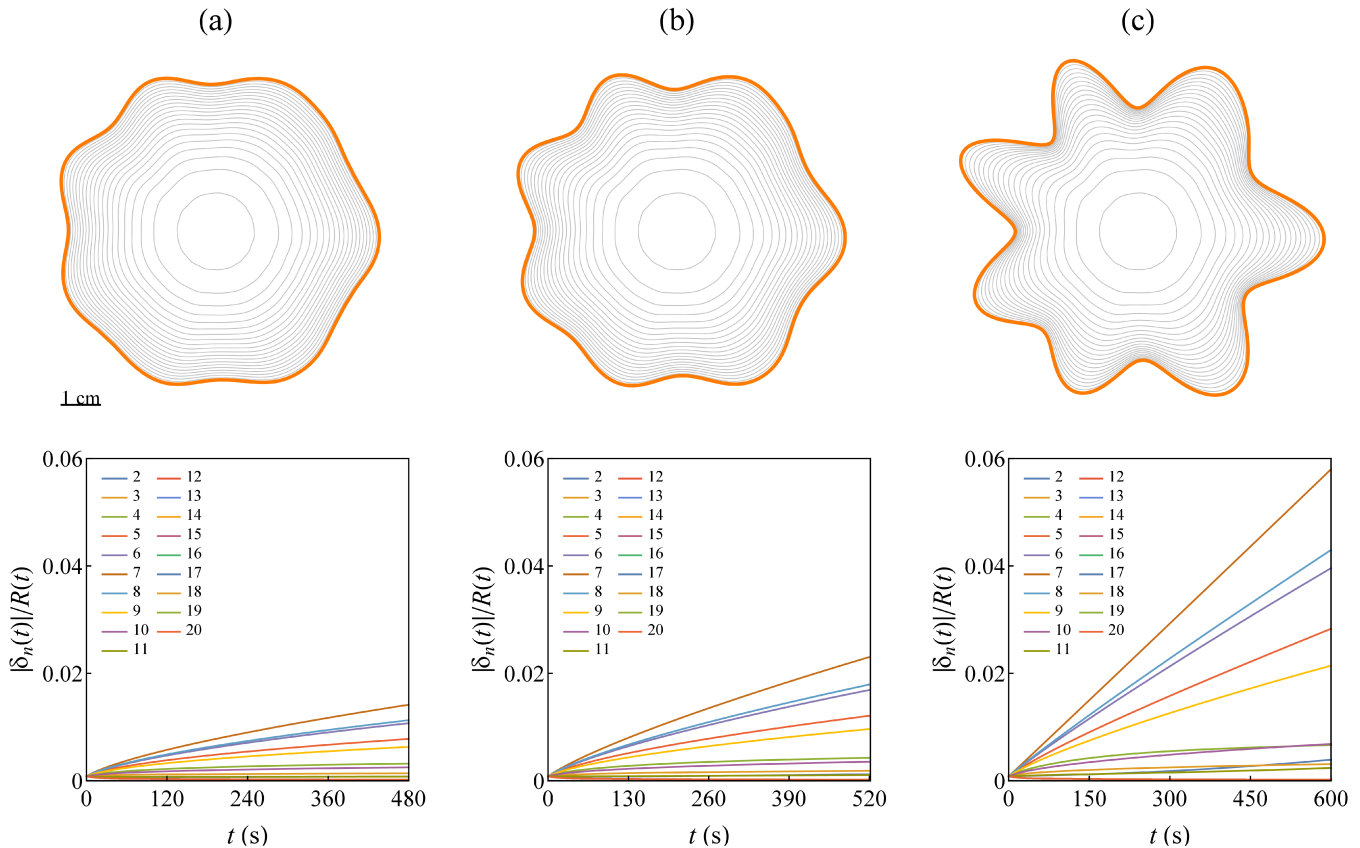} 
	
	\caption{{\bf Perturbative dynamics of expanding interfaces.} Weakly nonlinear time evolution of the expanding [$Q(t) > 0$] interfacial patterns generated by utilizing the controlling electric current $I(t)$ [Eq.~(\ref{Itime})] for $n_{\rm max}=7$, and three decreasing injection rates: (a) $Q(t)=0.25 \times Q_{\rm crit}(t)$, (b) $Q(t)=0.23 \times Q_{\rm crit}(t)$, and (c) $Q(t)=0.2 \times Q_{\rm crit}(t)$. Here we include modes $2 \leq n \leq 20$ and a random initial phase. The corresponding time evolution of the rescaled perturbation mode amplitudes $|\delta_{n} (t)|/R(t)$ are shown in the bottom panels. The final times used are (a) $t_f=480$ s, (b) $t_f=522$ s, and (c) $t_f=600$ s.}
	\label{WNL2}
\end{figure*}

Although the results depicted in Figs.~\ref{WNL2}-\ref{FNL2} are obtained utilizing weakly and fully nonlinear methods, our controlling protocol, the main result of this work, is derived from linear theory. Therefore, this section briefly describes the steps to obtain the linear growth rate of interfacial perturbations. Linear stability analysis of the problem~\cite{Bazant1} considers harmonic distortions of a nearly circular fluid-fluid interface whose position evolves according to ${\cal R}(\theta,t)= R(t) + \delta(\theta,t)$, where $R=R(t)$ is the time-dependent unperturbed radius of the interface, and $\theta$ denotes the azimuthal angle in the $r-\theta$ plane. The unperturbed radius of the interface at $t=0$ is denoted by $R(t=0)=R_0$, and the net interface disturbance is represented as a Fourier series $\delta(\theta,t)=\sum_{n=-\infty}^{+\infty} \delta_{n}(t)~e^{in\theta}$, where $\delta_{n}(t)$ denotes the complex Fourier amplitudes, with integer wave numbers $n$, and $|\delta| \ll R$. 
%\begin{equation}
%\label{fourier1}
%\delta(\theta,t)=\sum_{n=-\infty}^{+\infty} \delta_{n}(t)~e^{in\theta}, 
%\end{equation}
%where $\delta_{n}(t)$ denotes the complex Fourier amplitudes, with integer wave numbers $n$, and $|\delta| \ll R$. 

Recall that velocity and current density fields are irrotational in the bulk. Therefore, we can state our problem in terms of the Laplacian velocity and current density potentials~\cite{ST4,Bazant1,Bazant2} (see Supplementary Note 1). Within this perturbative framing, we use the kinematic boundary conditions~(\ref{kinematic1}) and~(\ref{kinematic2}) to express the potentials $\phi_{j}(t)$ and $\psi_{j}(t)$ in terms of the perturbation amplitudes $\delta_{n}(t)$, and their time derivatives ${\dot \delta_{n}}(t)= d \delta_{n}(t) / dt$. Next, we substitute the resulting relations, the pressure jump condition [Eq.~(\ref{pressure})], and continuity of electric potential [Eq.~(\ref{ectro})] into Darcy's law [Eq.~(\ref{Darcy1})]. Likewise, we also substitute these relations into the total current density [Eq.~(\ref{Darcy2})]. Then, since Eqs.~(\ref{Darcy1}) and~(\ref{Darcy2}) are coupled, we can insert one of them into the other, obtaining a single expression in terms of $\delta_{n}(t)$ and ${\dot \delta_{n}}(t)$. Finally, by keeping terms to first-order in $\delta$, and Fourier transforming, we obtain the equation of motion for the perturbation amplitudes $\dot{\delta}_{n}=\lambda(n) \delta_{n}$ (for $n \ne 0$), where
%\begin{eqnarray}
%	\label{result}
%	\dot{\zeta}_{n}&=&\lambda(n) \zeta_{n},
%\end{eqnarray}
\begin{eqnarray}
	\label{growth}
	\lambda(n)= \frac{Q}{2 \pi b R^{2}} (A |n| - 1) - \frac{\gamma}{R^{3}} B |n| (n^{2} - 1) + \frac{I}{\pi b R^2} C|n| \nonumber\\
\end{eqnarray}
is the linear growth rate, with electro-osmotic viscosity contrast
\begin{eqnarray}
\label{A}
A= \frac{K_{1}^2-K_{2}^2-(\sigma_{1}+\sigma_{2})(M_{1}-M_{2})}{(K_{1}+K_{2})^2-(\sigma_{1}+\sigma_{2})(M_{1}+M_{2})}.
\end{eqnarray}
Additionally,
\begin{eqnarray}
\label{B}
B= \frac{M_{1}K_{2}^2 + M_{2}K_{1}^2-(\sigma_{1}+\sigma_{2})M_{1}M_{2}}{(K_{1}+K_{2})^2-(\sigma_{1}+\sigma_{2})(M_{1}+M_{2})},
\end{eqnarray}
and
\begin{eqnarray}
\label{C}
C= \frac{M_{1}K_{2} - M_{2}K_{1}}{(K_{1}+K_{2})^2-(\sigma_{1}+\sigma_{2})(M_{1}+M_{2})}.
\end{eqnarray}
The first term appearing on the right-hand side of Eq.~(\ref{growth}) represents the destabilizing contribution (for $A>0$) coming from the radial injection of the inner fluid, while the second term accounts for the stabilizing effect due to the surface tension. Supplementing these usual terms related to injection-driven Hele-Shaw flows, there is also an additional term proportional to $IC$, which arises as a consequence of the consideration of the electro-osmotic flow. This new term can promote linear stabilization ($IC < 0$) or destabilization ($IC > 0$) depending on its sign. In addition, note that at the linear level electro-osmotic effects also modify the parameters $A$ and $B$, which, in the absence of electro-osmotic effects ($K_{1}=K_{2}=0$), are given by $A=(\eta_{2}-\eta_{1})/(\eta_{2}+\eta_{1})$ and $B=b^2/[12(\eta_{1}+\eta_{2})]$, respectively. In Supplementary Note 1, we extend the linear theory to a second-order mode-coupling weakly nonlinear analysis.

\subsection*{Controlling protocol}
\label{discuss}

Our goal in this section is to design a controlling scheme focused on prescribing the number of
fingers emerging at the fluid-fluid interface by properly adjusting the flow injection rate and electric current. In contrast to what has been employed in Refs.~\cite{Bazant1,Bazant2}, instead of using usual constant values for $Q$ and $I$, which may result in finger proliferation, here we consider that both quantities may be functions of time, i.e., $Q=Q(t)$ and $I=I(t)$. At the linear level, an estimate for the number of fingers formed during the injection process is given by the closest integer to the mode of largest growth rate $n_{\rm max}$~\cite{Lp,Rev1,Rev2,Rev3}, found by evaluating $d\lambda/dn|_{n=n_{\rm max}}=0$. Since we intend to keep this number fixed, our job is to determine the functional forms of $Q(t)$ and $I(t)$ that keep $n_{\rm max}$ unmodified as the interface evolves. Considering Eq.~(\ref{growth}), $d\lambda/dn|_{n=n_{\rm max}}=0$ is easily evaluated. Then, by solving the resulting expression for $I(t)$, we obtain
%leading to 
%\begin{eqnarray}
%\label{nmax}
%I(t)C+\frac{Q(t)A}{2}-\frac{\pi b \gamma B}{R(t)}(3{n_{\rm max}^2}-1)=0.
%\end{eqnarray}
\begin{eqnarray}
\label{Itime}
I(t)=\frac{1}{C}\left[\frac{\pi b \gamma B}{R(t)}(3{n_{\rm max}^2}-1) - \frac{Q(t)A}{2} \right].
\end{eqnarray}
This is the required time-dependent electric current needed to maintain the number of fingers fixed during the injection-driven, electro-osmotic flow in a radial Hele-Shaw cell. Expression~(\ref{Itime}) constitutes one of the central analytical results of this work. Note that it does not impose any restrictions on the functional form and values of the injection rate $Q(t)$. Therefore, {\it a priori}, any $Q(t)$, including constant values, could be utilized to keep $n_{\rm max}$ unmodified as long as $I(t)$ varies in time accordingly to Eq.~(\ref{Itime}). However, this is not true, and we address this point in the following calculations.

By inserting Eq.~(\ref{Itime}) into Eq.~(\ref{growth}), and evaluating the resulting expression at $n=n_{\rm max}$, we obtain the growth rate of mode $n_{\rm max}$
\begin{eqnarray}
\label{lamb2}
\lambda(n_{\rm max})= \frac{1}{R^3(t)}\left[2  \gamma B {n^3_{\rm max}} - \frac{Q(t)R(t)}{2 \pi b}\right].
\end{eqnarray}
Note that if one assumes that the injection process is performed under constant rate, i.e., $Q(t)=Q_{0}$, the second term inside the squared brackets in Eq.~(\ref{lamb2}) dominates for larger interfacial sizes, and $n_{\rm max}$ becomes a stable mode [$\lambda(n_{\rm max})<0$]. Recall that $n_{\rm max}$ is the mode of largest growth rate, so when it becomes a stable mode, all the other modes in the dynamics become stable as well. Consequently, the interface expands as a stable circle and not as a controlled boundary with a specific number of fingers, as we would like. Therefore, to control the shape of the interface during injection-driven, electric-osmotic flows, one must consider not only a time-dependent electric current [Eq.~(\ref{Itime})], but also a time-dependent injection rate. Moreover, in order to guarantee that $n_{\rm max}$ is an unstable mode during all the injection process, and to ensure the formation of fingered structures, we need that $\lambda(n_{\rm max})>0$, leading to the requirement
\begin{eqnarray}
\label{requir}
Q(t) < Q_{\rm crit} (t) = \frac{4 \pi b \gamma B {n^3_{\rm max}}}{R(t)},
\end{eqnarray}
where $Q_{\rm crit} (t)$ is the critical injection rate at which $n_{\rm max}$ becomes stable [i.e., $\lambda(n_{\rm max})=0$].

Based on the previous calculations and discussions, it is clear that if one intends to control the shape of the expanding interface during injection-driven, electric-osmotic flows, one needs to consider the controlling electric current~(\ref{Itime}) and perform the injection process under a positive, time-dependent, non-zero injection rate $Q(t)=\alpha Q_{\rm crit}(t)$, where $0<\alpha<1$. Since $Q(t)=2\pi b R \dot R$, $R(t)$ can be easily determined, and once this is done one concludes that $Q(t) \sim t^{-1/3}$. 

We also point to the fact that only the controlling electric current~(\ref{Itime}) depends on $n_{\rm max}$, and on the physical parameters of the system, while $Q(t)$ can be arbitrarily chosen [as long as condition~(\ref{requir}) is satisfied] to control other features of the interface and tune dynamical regimes, without the necessity of changing the set of physical parameters, initial conditions or number of fingers. We stress that these controlling features are not possible in the purely hydrodynamic problem. When electric-osmotic effects are absent, i.e., $I(t)=0$ and $K_1=K_2=0$, Eq.~(\ref{Itime}) reduces to $Q(t) \equiv Q_{\rm ph}(t) =2 \pi b \gamma B (3n^3_{\rm max}-1)/A R(t)$, which is the controlling injection rate for the pure hydrodynamic problem. In this case, if one intends to induce dynamical responses of the interface by manipulating the injection $Q_{\rm ph}(t)$ while keeping the number of fingers fixed ($n_{\rm max}$), then one necessarily has to modify the physical parameters. Likewise, if the physical parameters are kept unaltered but $Q_{\rm ph}(t)$ is adjusted to tune dynamical behaviors, then $n_{\rm max}$ has to change, impacting the interfacial symmetry. All of these problems are eliminated by our new controlling protocol. Therefore, according to the linear theory, the inclusion of electric-osmotic effects in an injection-driven Hele-Shaw flow allows one to obtain a family of interfacial patterns, all of them with the same number of fingers emerging at their boundaries [set by the value of $n_{\rm max}$ and prescribed by $I(t)$], but generated employing different injection rates $Q(t)$.

%Once the physical parameters and initial conditions are determined, and the desired number of fingers is chosen (given by the integer value of $n_{\rm max}$), there is only one pattern associated with $ Q_{\rm ph}(t)$, in contrast to the family of fingered structures that arises when the electric-osmotic flow is also considered.

In the remainder of this paper, we go beyond the linear regime to test the efficiency of our linear-stability-based strategy [Eqs.~(\ref{Itime}) and~(\ref{requir})] in controlling the number of fingers emerging at the interface and investigate the impact of different injection rates on the dynamic responses and morphological aspects of the expanding interfaces in weakly and fully nonlinear regimes of the dynamics. To strengthen the relevance of our theoretical results, the values of the parameters used in our simulations (both weakly and fully nonlinear) are based on the values of the physical quantities utilized in the experimental paper~\cite{Bazant2} related to injection-driven, electro-osmotic flows in a radial Hele-Shaw cell.

\subsection*{Nonlinear dynamics of the expanding interfaces}
\label{self}

In this section, we consider an injection of oil into a radial Hele-Shaw cell filled with a mixture of water and glycerol. In such circumstances, and considering the physical parameters listed in the section Methods, $A \approx 0.87, B \approx 7.16 \times 10^{-9}$ m$^4~\cdot$ N$^{-1}\cdot$ s$^{-1}$ $ > 0$, and $C \approx -1.80 \times 10^{-8}$ m$^3~\cdot$ A$^{-1}\cdot$ s$^{-1}$$< 0$. Therefore, according to Eq.~(\ref{growth}), injection always acts to promote interfacial instabilities, while positive (negative) electric currents have stabilizing (destabilizing) effects on the interface.

In order to verify the efficiency of the controlling protocol during the onset of nonlinearities, we first extend the linear theory to a second-order perturbative approach of the problem and utilize it to describe the weakly nonlinear evolution of the interface (see Supplementary Note 1). Then, we consider the nonlinear coupling of sine and cosine Fourier modes in the interval $2 \leq n \leq 20$, and obtain the time evolution of the mode amplitudes by numerically solving the corresponding coupled nonlinear differential equations. Moreover, we set the initial amplitude of all perturbation modes as $R_{0}/600$, and in order to make the initial conditions as general as possible, we consider the action of random phases attributed to each participating sine and cosine mode. 

In the top panels of Fig.~\ref{WNL2} we plot the weakly nonlinear evolution of the interface employing the controlling current~(\ref{Itime}) for $n_{\rm max}=7$ and three decreasing injection rates: (a) $Q(t)=0.25 \times Q_{\rm crit}(t)$, (b) $Q(t)=0.23 \times Q_{\rm crit}(t)$, and (c) $Q(t)=0.2 \times Q_{\rm crit}(t)$. When plotting these temporal evolutions, time varies in the range $0\leq t \leq t_{f}$, where the final time $t_f$ is defined as the time at which the interface unperturbed radius has reached the same magnitude [namely, $R(t = t_f) \approx 4$ cm] for each $Q(t)$ employed. This is done with no loss of generality, in order to make the generated patterns to have the same size at $t = t_f$. In these circumstances, the final times are taken as (a) $t_f=480$ s, (b) $t_f=522$ s, and (c) $t_f=600$ s. In the bottom panels we show the corresponding time evolution of the rescaled perturbation amplitudes $|\delta_{n} (t)|/R(t)$.

\begin{figure*}[t]
	\centering
	\includegraphics[scale=0.95]{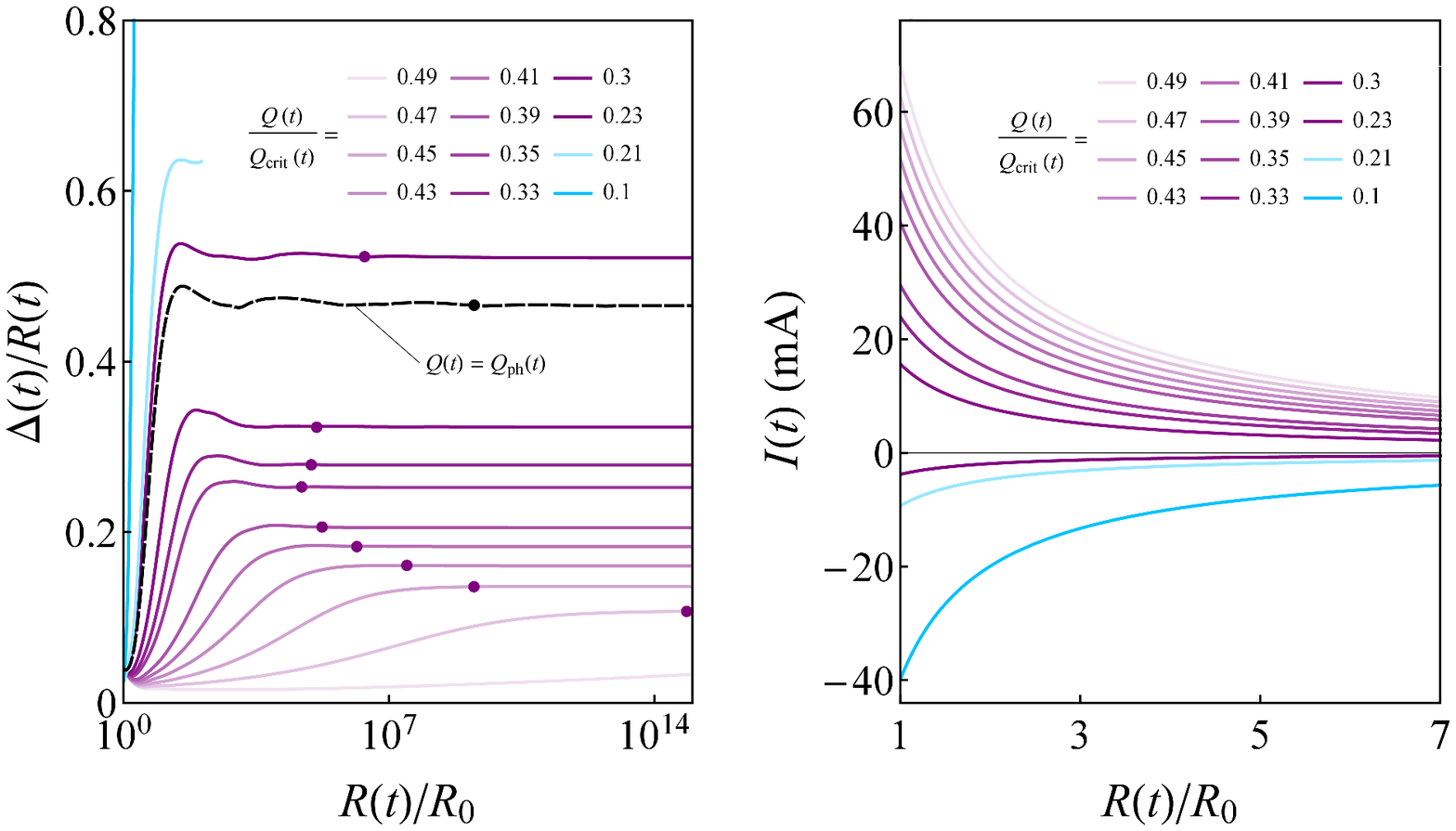} 
		
	\caption{{\bf Analysis of the self-similar growth.} Left panel: Behavior of the shape factor $\Delta(t)/R(t)$ with respect to variations in the ratio $R(t)/R_0$, for flows performed utilizing the electric current~(\ref{Itime}) with $n_{\rm max}=7$ and various injection rates $Q(t)$. The small dots indicate the points in which the shape factor saturates, i.e., the onset of the self-similar regime. For the curve associated with $Q(t)=0.49 \times Q_{\rm crit}(t)$, the small dot lies outside the plot interval. As for the curves related to $Q(t)=0.21 \times Q_{\rm crit}(t)$ and $Q(t)=0.1 \times Q_{\rm crit}(t)$, these cases do not reach the self-similar evolution. Right panel: Plot of the electric current~(\ref{Itime}) as a function of $R(t)/R_0$ for the injection rates utilized in the left panel.}
	\label{shapefac}
\end{figure*}

By scrutinizing the top panels of Fig.~\ref{WNL2}, we observe that all the patterns evolve to 7-fold fingered structures, regardless of the injection rate employed. These three interfacial shapes are just a few examples of the many structures composing the family of 7-fold patterns that can be found by varying the injection rate. These results are in very good agreement with the predictions of the linear theory. In addition, note that $Q(t)$ has a peculiar effect on the interfacial patterns: Although all of them have the same number of fingers and are generated employing precisely the same initial conditions and physical parameters, it seems that lower injection rates lead to the formation of increasingly disturbed patterns. This is very clear when one contrasts the interfacial shape depicted in Fig.~\ref{WNL2}(c) to the structures shown in Figs.~\ref{WNL2}(a) and~\ref{WNL2}(b). As a matter of fact, this effect can also be perceived and appropriately quantified by noting that the magnitudes of the rescaled perturbation amplitudes $|\delta_{n} (t_f)|/R(t_f)$ in the bottom panel of Fig.~\ref{WNL2}(c) are larger than the magnitudes found in the bottom panels of Figs.~\ref{WNL2}(a) and~\ref{WNL2}(b). Fig.~\ref{WNL2} unveils the first very welcome role of $Q(t)$ in our controlling protocol, namely the ability to tune the level of instability of the $n$-fold morphologies. 

\begin{figure*}[t]
	\centering
	\includegraphics[scale=0.62]{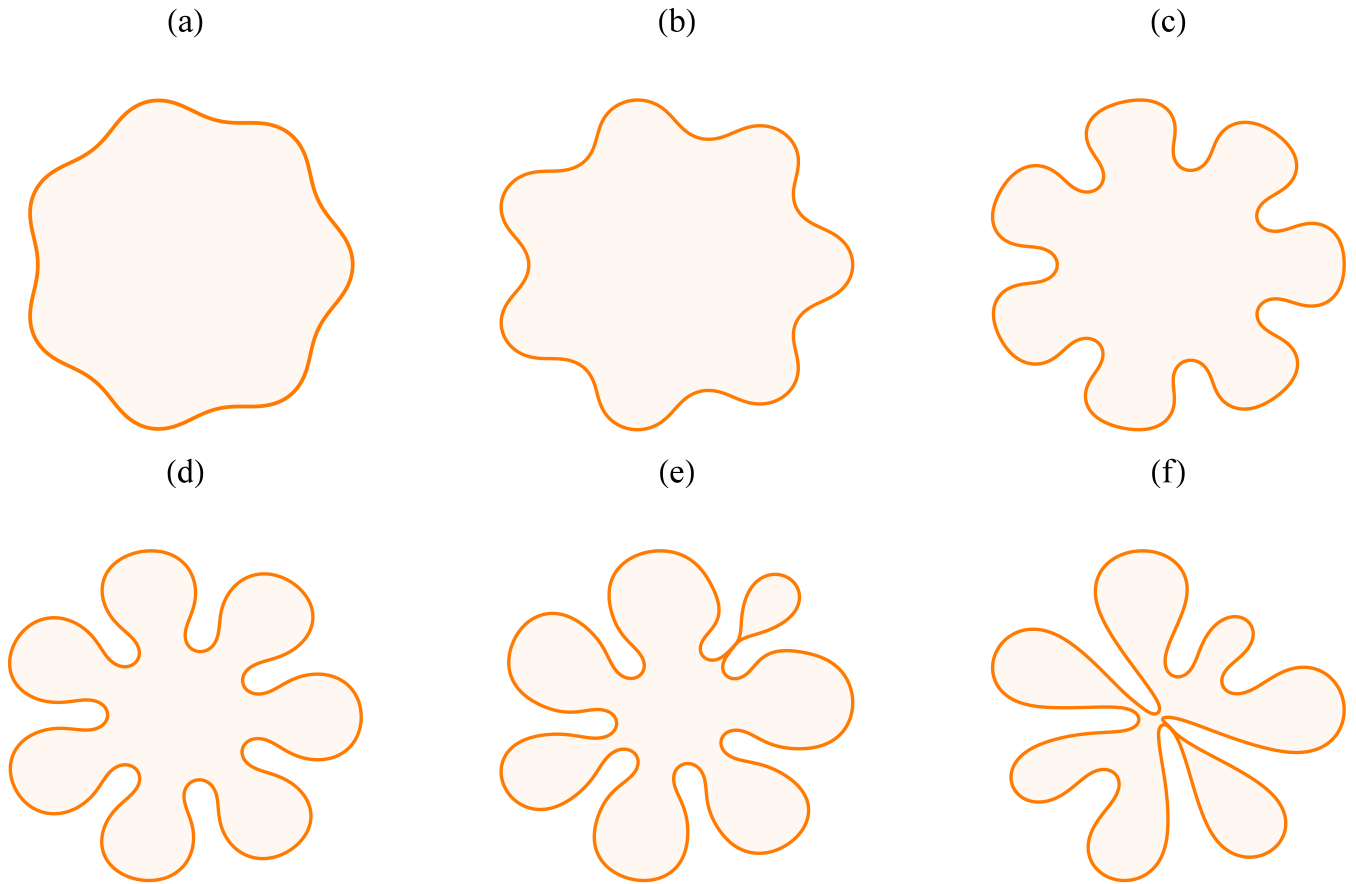} 
	
	\caption{{\bf Snapshots of fully nonlinear interfaces for different injection rates.} Gallery of representative fully nonlinear 7-fold patterns corresponding to the cases (a) $Q(t)=0.49 \times Q_{\rm crit}(t)$, (b) $Q(t)=0.45 \times Q_{\rm crit}(t)$, (c) $Q(t)=0.3 \times Q_{\rm crit}(t)$, (d) $Q(t)=0.23 \times Q_{\rm crit}(t)$, (e) $Q(t)=0.21 \times Q_{\rm crit}(t)$, and (f) $Q(t)=0.1 \times Q_{\rm crit}(t)$, shown in Fig.~\ref{shapefac}. The interfaces (a)-(d) are plotted at the onset of the self-similar regime, i.e., for (a) $R/R_{0} = 10^{36}$, (b) $R/R_{0} = 1.75 \times 10^{9}$, (c) $R/R_{0} = 1.25 \times 10^{5}$, and (d) $R/R_{0} = 2.29 \times 10^{6}$. The patterns depicted in (e) and (f) do not reach the self-similar regime, and are plotted for (e) $R/R_{0} = 114$ and (f) $R/R_{0} = 1.97$. The videos of the evolution of the interfaces (c)-(f) are provided in Supplementary Movies 1-4, respectively.}
	\label{FNL2}
\end{figure*}

In the framework of pure hydrodynamic, injection-driven radial Hele-Shaw flows, it has been demonstrated by fully nonlinear simulations~\cite{ShuwangPRL,pedrowet} that flows performed under time-dependent injection fluxes scaling as $Q(t)\sim t^{-1/3}$, such as the injection rates utilized in this work, exhibit formation of well behaved $n$-fold symmetric, self-similar patterns. These interesting fingered morphologies are formed at very long times of the dynamics in a highly nonlinear regime characterized by the emergence of radially evolving structures with preserved shapes. Therefore, before reaching the self-similar regime, the patterns evolve as structures with a fixed number of fingers, though their shapes are continuously changing. After reaching the self-similar regime, the interface still expands radially but with unaltered morphology. To investigate whether or not our controlling scheme, based on the usage of Eqs.~(\ref{Itime}) and~(\ref{requir}), is still efficient in keeping the number of fingers fixed in the fully nonlinear regime, and to study the effects (if any) of different injection rates $Q(t)$ in the establishment of the self-similar regime, and in the associated pattern morphologies, we resort to our boundary integral method. The numerical algorithm utilized here extends the boundary integral codes previously reported in Refs.~\cite{ShuwangJCP,Zhao17} to include the contributions coming from the applied electric field, as we explain in detail in Supplementary Note 2.

%This type of evolution is illustrated in Fig.~\ref{WNL2}.

A convenient way to investigate the intrinsically fully nonlinear concerns mentioned in the previous paragraph is through the time evolution of the interfacial shape factor $\Delta(t) / R(t)$~\cite{ShuwangPRL}. The shape factor $\Delta(t) / R(t) = {\rm max}|{\cal R}(\theta,t)/R(t)-1|$ is computed numerically based on the maximum deviation of the perturbed interface from the equivalent circle with the same area. With the application of the electric field, supplemented by the injection process, the interface deforms, and in the course of time, its shape factor starts to grow. If the radially expanding pattern eventually reaches a stable state of permanent profile, its shape factor ceases to vary with time [or, equivalently, with $R(t)$], and the self-similar regime is achieved. On the other hand, if the pattern keeps changing its morphology while evolving, its shape factor will keep varying as time progresses, and the establishment of self-similar regime does not occur.

In the left panel of Fig.~\ref{shapefac}, we present the behavior of the shape factor $\Delta(t) / R(t)$ with respect to variations in the ratio $R(t)/R_{0}$ for thirteen different injection rates. The solid curves correspond to injection-driven, electro-osmotic flows performed utilizing the controlling current~(\ref{Itime}) with $n_{\rm max}=7$ and injection rates $Q(t)=\alpha Q_{\rm crit}(t)$. On the other hand, the dashed curve represents the pure hydrodynamic flow performed using $Q(t)=Q_{\rm ph}(t)$. We stress that this graph is plotted by considering the evolution of the patterns given by a boundary integral method and not by the perturbative approach utilized in Fig.~\ref{WNL2}. In addition, in the right panel of Fig.~\ref{shapefac} we depict the electric current~(\ref{Itime}) as a function of $R(t)/R_{0}$ for the injection rates utilized in the left panel.

By following the behavior of the curves in Fig.~\ref{shapefac}, initially, one observes the growth of $\Delta(t) / R(t)$, indicating the regime in which the interface changes its morphology while it expands radially. Note that for a given value of $R(t)/R_{0}$, this growth is steeper for lower injection rates, which is in agreement with the weakly nonlinear results shown in Fig.~\ref{WNL2} regarding the role of the injection rate on regulating the pattern's instability. Nonetheless, as time advances and the interface acquires large sizes, fully nonlinear effects dominate, making most of the curves saturate. Small dots indicate the moment of saturation. From these points onward, the corresponding interfaces evolve self-similarly, keeping a constant interfacial shape profile. We emphasize that this does not mean that the system reaches a stationary state since the interfaces still expanding radially. Note that although the curve associated to $Q(t)=0.49 \times Q_{\rm crit}(t)$ reaches the self-similar regime, it occurs for a very large interfacial size ($R/R_{0}= 10^{36}$) and the small dot lies outside the plot interval. On the other hand, the cases with $Q(t)=0.1 \times Q_{\rm crit}(t)$ and $Q(t)=0.21 \times Q_{\rm crit}(t)$ do not achieve the self-similar regime, and the reason behind this failure in reaching such regime will be provided later when analyzing Fig.~\ref{FNL2}.

However, the most noteworthy dynamical behavior revealed by Fig.~\ref{shapefac} is the possibility of tuning the occurrence of the self-similar regime by varying the injection rate. We observe that by properly manipulating the injection rate $Q(t)$ and the electric current accordingly to Eq.~(\ref{Itime}), one can tune the establishment of the self-similar regime to occur at interfacial sizes as large as $R/R_{0}= 10^{36}$ [for $Q(t)=0.49 \times Q_{\rm crit}(t)$], or as small as $R/R_{0}= 5.02 \times 10^4$ [for $Q(t)=0.35 \times Q_{\rm crit}(t)$]. In addition, by utilizing $Q(t)/Q_{\rm crit}(t) \leq 0.21$ one can even stop the occurrence of self-similarity.  These novel controlling features are only possible due to the inclusion of the electro-osmotic flow in addition to the usual pressure-driven flow. To make this point very clear, in Fig.~\ref{shapefac} we also depict the evolution of the shape factor during a purely hydrodynamic flow performed employing $Q(t)=Q_{\rm ph}(t)$. The dashed curve represents this situation. Although in this case, one also observes establishment of self-similar evolution at $R/R_{0}= 1.77 \times 10^9$, there are no ways of tuning its occurrence (delaying, promoting, or preventing) without modifying $n_{\rm max}$, the physical parameters, or the initial conditions of the flow. Conversely, our controlling protocol based on the usage of applied electric fields permits control of all these features while keeping all the physical parameters, initial conditions, and $n_{\rm max}$ unaltered. In particular, we point out that, comparing the curve related to $Q(t)=0.35 \times Q_{\rm crit}(t)$ with the one associated to $Q(t)=Q_{\rm ph}(t)$, our controlling scheme has promoted the establishment of self-similar evolution by reducing the required interfacial size $R/R_{0}$ in about $5$ orders of magnitude with respect to the equivalent purely hydrodynamic case.

The analysis of Fig.~\ref{shapefac} revealed an interesting fully nonlinear dynamical response of the system to the injection rates employed, namely the significantly reduction (increase) of the interfacial size $R/R_{0}$ associated with the onset of the self-similar evolution when lower (higher) injection rates are utilized. However, the reduction trend is verified up until $Q(t)/Q_{\rm crit}(t)=0.35$. Below that value, we verified the opposite behavior, i.e., an increasing in $R/R_{0}$ when injection rates $Q(t)/Q_{\rm crit}(t) < 0.35$ are employed, and complete absence of self-similarity for $Q(t)/Q_{\rm crit}(t) \leq 0.21$. Complementary information about this interruption in the establishment of self-similar growth is provided in Fig.~\ref{FNL2}, where we plot the fully nonlinear patterns related to some of the cases presented in Fig.~\ref{shapefac}. More specifically, we depict the interfacial shapes for the following values of $Q(t)/Q_{\rm crit}(t)$: $0.49$ [Fig.~\ref{FNL2}(a)], $0.45$ [Fig.~\ref{FNL2}(b)], $0.3$ [Fig.~\ref{FNL2}(c)], $0.23$ [Fig.~\ref{FNL2}(d)],  $0.21$ [Fig.~\ref{FNL2}(e)], and $0.1$ [Fig.~\ref{FNL2}(f)]. The patterns (a)-(d) are displayed at the onset of the self-similar regime, for (a) $R/R_{0} = 10^{36}$, (b) $R/R_{0} = 1.75 \times 10^{9}$, (c) $R/R_{0} = 1.25 \times 10^{5}$, and (d) $R/R_{0} = 2.29 \times 10^{6}$. On the other hand, the non self-similar fingered structures (e) and (f) are plotted for $R/R_{0} = 114$ and $R/R_{0} = 1.97$, respectively.

By inspecting the panels of Fig.~\ref{FNL2}, one immediately notes that all the fingered structures are 7-fold, confirming the efficiency of our method to control the number of emerging fingers regardless of the injection rate employed. Moreover, it is evident that lower injection rates indeed lead to the formation of more disturbed patterns. These findings are in line with the previous weakly and fully nonlinear results presented in Figs.~\ref{WNL2} and~\ref{shapefac}, respectively. Nevertheless, by reducing $Q(t)/Q_{\rm crit}(t)$, we observe a morphological transition from 7-fold, symmetric, self-similar patterns [panels (a)-(d)], to asymmetric 7-finger structures [panels (e) and (f)]. This is clearly illustrated in Fig.~\ref{FNL2}(e), where we observe the formation of an almost symmetric pattern that seems to evolve towards self-similarity, but this evolution breaks due to the occurrence of a pinch-off event. Finally, in Fig.~\ref{FNL2}(f), we observe the emergence of an asymmetric, intricate fingered shape, dominated by electric-induced pinch-off instabilities with no resemblance to the well-behaved symmetric patterns depicted in the other panels for larger values of $Q(t)/Q_{\rm crit}(t)$. This morphological transition is explained by inspecting the right panel of Fig.~\ref{shapefac}: As $Q(t)/Q_{\rm crit}(t)$ is reduced, one diminishes the destabilizing viscous effects but also restrain the stabilization provided by positive electric currents. When the electric current crosses the $x$-axis and thus becomes negative, the electric field turns a destabilizing effect, ultimately leading to pinch-off phenomena for sufficiently large currents. Supplementary Movies 1-4 present the evolution of the patterns displayed in Figs.~\ref{FNL2}(c)-(f), respectively. In Supplementary Note 3, we analyze the weakly and fully nonlinear pattern-forming dynamics in the absence of injection [$Q(t)=0$]. In addition, Supplementary Note 4 shows the uncontrolled, disordered growth of the interface when constant values of $I$ and $Q$ are employed. And finally, in Supplementary Notes 5 and 6, we show that the dynamical behaviors found in our work are general, in the sense that if one selects other modes $n_{\rm max}$, the key conclusions remain qualitatively the same.

\section*{Discussion}

Above we show through numerical simulations that dynamical control of fingering instabilities is attained by properly adjusting both electric current and flow rate over time. Our controlling protocol differs from previous strategies performed under time-varying flow rates because of the inclusion of a secondary electro-osmotic flow, which can oppose or assist the pressure-driven flow. Remarkably, this simple modification has provided an improved control over the features and behaviors of the unstable viscous fingering interface. More specifically, we are able to set the number of fingers emerging at the interface, tune the instabilities' level and pattern morphology, and control if and when self-similar growth occurs. We stress that all these features are conveniently controlled without altering the material properties of the fluids, physical parameters, or initial conditions, something impossible to be realized by utilizing conventional strategies performed solely in terms of time-varying flow rates~\cite{ShuwangPRL,mama1,irio,woods,pedrowet}.

Many extensions of our work are possible. Although we demonstrate improved interfacial control by utilizing electric fields, our method is general and can be applied to many other systems where multiple forces drive interfacial instabilities. For instance, one could use our discoveries to control the features of a radially evolving interface by applying magnetic fields instead of electric fields. In that case, the fluids are magnetic and respond promptly to applied magnetic fields~\cite{pedromag,lab1,lab2,lab3,lab4,lab5,Chen1,Chen2,rad1,rad2,crist}. Furthermore, given the existing types of magnetic fluids (ferrofluids and magnetorheological)~\cite{pedromag} and the different possibilities of magnetic field arrangements~\cite{pedromag,lab1,lab2,lab3,lab4,lab5,Chen1,Chen2,rad1,rad2,crist}, these systems could potentially exhibit new dynamical behaviors beyond the ones already disclosed here when controlled by employing our protocol.

In conclusion, we have reported a novel controlling strategy performed under the employment of time-dependent electric currents and flow rates. Our ``non-invasive" method provides improved control over features of an unstable interface during radial Hele-Shaw flow and triggers dynamical behaviors spanning at the fully nonlinear regime of the dynamics by only manipulating external driven forces while keeping physical parameters and fluid properties unaltered. A key factor in achieving such control was the consideration of an additional flow (electro-osmotic). Therefore, our present study paves the way for other explorations concerning controlling methods exploiting the rich physics behind ``multi-field" driven interfacial dynamics. Our numerical findings substantiate our linear and weakly nonlinear stability predictions, supporting the pertinence and usefulness of the proposed controlling process at nonlinear stages. Thus, we have designed a superior strategy that provides a step forward from the current research on controllability toward ultimate control of complex viscous fingering patterns. Finally, our results are also relevant to many fluid dynamics systems in confined geometries and microfluidic devices, in which it is desirable to be able to prescribe the shape and symmetry properties of viscous fingering patterns, creating the possibility of designing materials uniquely targeted to specific applications. 

\section*{Methods}

\subsection*{Physical parameters and initial conditions}

In agreement with the physical parameters utilized to perform the experiments reported in Ref.~\cite{Bazant2}, the results obtained in our work [Figs.~\ref{WNL2}-\ref{FNL2}] consider the injection of oil with ${\eta}_1 = 7.36 \times 10^{-3}$ Pa s, $\varepsilon_{1} = 10.3 \varepsilon_{0}$, and $\zeta_{1} = 0$ V, into a radial Hele-Shaw cell filled with a mixture of water and glycerol with ${\eta}_2 = 109 \times 10^{-3}$ Pa s, $\varepsilon_{2} = 49.1 \varepsilon_{0}$, and $\zeta_{2} = -150 \times 10^{-3}$ V, where $\varepsilon_{0} \approx 8.85 \times 10^{-12}$ F/m is the vacuum permittivity. In addition, $\gamma = 37 \times 10^{-3}$ N/m, $\sigma_{1} = \sigma_{2} = 155 \times 10^{-4}$ S/m, $b=10^{-4}$ m, and $R_0=10^{-2}$ m. For the initial conditions, please, see Supplementary Notes 1 and 2.

\subsection*{Weakly nonlinear plots}

See Supplementary Note 1.

\subsection*{Boundary integral simulations}

See Supplementary Note 2.

\section*{Data availability}

The data supporting the findings of this study are available within the paper and its supplementary files, and are available from the corresponding author upon request.

\begin{acknowledgments}
	
P. A. acknowledges useful discussion with Jos\'e Miranda, Eduardo Dias, and \'Irio Coutinho. S. L. acknowledges the support from the National Science Foundation, Division of Mathematical Sciences grant DMS-1720420. J. L. acknowledges partial support from the NSF through grants DMS-1714973, DMS-1719960, DMS-1763272, and the Simons Foundation (594598QN) for a NSF-Simons Center for Multiscale Cell Fate Research. J. L. also thanks the National Institutes of Health for partial support through grants 1U54CA217378-01A1 for a National Center in Cancer Systems Biology at UC Irvine and P30CA062203 for the Chao Family Comprehensive Cancer Center at UC Irvine.
	
\end{acknowledgments}

\section*{Author contributions}

P. A. designed the controlling protocol, performed linear and weakly nonlinear theories, and drafted the manuscript. M. Z. performed the numerical simulations. S. L. initiated the project. P. A., M. Z., J. L., and S. L. discussed and interpreted the results.

\section*{Competing interests}

The authors declare no competing interests.

\end{document}

\title{Supplementary information: \\ Electrically-controlled self-similar evolution of viscous fingering patterns in radial Hele-Shaw flows}
\author{Pedro H. A. Anjos$^{1}$}
\email[]{pamorimanjos@iit.edu (corresponding author)}
\author{Meng Zhao$^{2}$}
\email[]{mzhao9@hust.edu.cn}
\author{John Lowengrub$^{3}$}
\email[]{lowengrb@math.uci.edu}
\author{Shuwang Li$^{1}$}
\email[]{sli@math.iit.edu}
\affiliation{$^{1}$Department of Applied Mathematics, Illinois Institute of Technology,
	Chicago, Illinois 60616, USA \\ 
	$^2$ Center for Mathematical Sciences, Huazhong University of Science and Technology, Wuhan 430074, China \\
$^3$Department of Mathematics, University of California at Irvine, Irvine, California 92697, USA}

\maketitle

\def\blankpage{%
	\clearpage%
	\thispagestyle{empty}%
	\addtocounter{page}{-1}%
	\null%
	\clearpage}

\section*{Supplementary Note 1: Perturbative weakly nonlinear approach}
\label{WNL}

In this section, we develop a second-order, perturbative weakly nonlinear theory for our electrohydrodynamic problem. During the injection process, the initially slightly perturbed circular interface can become unstable, and deform, due to the interplay of viscous, capillary, and electric forces acting on the system. In our perturbative analysis, the perturbed shape of the fluid-fluid boundary is described by ${\cal R}(\theta,t)= R(t) + \delta(\theta,t)$, where $R=R(t)$ is the time-dependent unperturbed radius of the interface, and $\theta$ denotes the azimuthal angle in the $r-\theta$ plane. The net interface disturbance is represented as a Fourier series
\begin{equation}
\label{fourier1}
\delta(\theta,t)=\sum_{n=-\infty}^{+\infty} \delta_{n}(t)~e^{in\theta}, 
\end{equation}
where $\delta_{n}(t)$ denotes the complex Fourier amplitudes, with integer wave numbers $n$, and $|\delta| \ll R$. Mass conservation imposes that the zeroth mode is written in terms of the other modes as $\delta_{0}= -(1/2R) \sum_{n=1}^{\infty} \left [|\delta_{n}(t)|^2 + |\delta_{-n}(t)|^2 \right ]$~\cite{ST4}.

Recall that the velocity ${\textbf v}_{j}$ and current density ${\textbf J}_{j}$ fields are irrotational in the bulk. Therefore, we can state our problem in terms of the Laplacian velocity and current density potentials~\cite{ST4,Bazant1,Bazant2}
\begin{eqnarray}
\label{phi} 
\phi_{j}(r,\theta) =- \frac{Q}{2\pi b} \log \left(\frac{r}{R} \right) 
+ \sum_{n \ne 0}~\phi_{j n}(t)~\left(\frac{r}{R}\right)^{(-1)^{(j+1)} |n|}~e^{in\theta},
\end{eqnarray}
and
\begin{eqnarray}
\label{psi} 
\psi_{j}(r,\theta) =- \frac{I}{2\pi b} \log \left(\frac{r}{R} \right) 
+ \sum_{n \ne 0}~\psi_{j n}(t)~\left(\frac{r}{R}\right)^{(-1)^{(j+1)} |n|}~e^{in\theta},
\end{eqnarray}
respectively. Here our leading goal is to find a differential equation which describes the time evolution of the perturbation amplitudes $\delta_{n}(t)$, accurate to second-order [$O(\delta_{n}^{2})$]. Therefore, we perform the same steps already utilized to obtain the linear growth rate of the system, but instead of keeping terms up to first-order in $\delta$, we extend the linear theory by keeping terms consistently up to second-order in $\delta$. By doing this, we obtain the {\it weakly nonlinear} equation of motion for the perturbation amplitudes (for $n \ne 0$)
\begin{eqnarray}
	\label{result}
	\dot{\delta}_{n}=\lambda(n) \delta_{n} + \sum_{n' \neq 0} \left [ F(n, n')  + \lambda(n') G(n, n')  \right ]\delta_{n'} \delta_{n - n'}. 
\end{eqnarray}
In Eq.~(\ref{result}), the function 
\begin{eqnarray}
	\label{growth}
	\lambda(n)= \frac{Q}{2 \pi b R^{2}} (A |n| - 1) - \frac{\gamma}{R^{3}} B |n| (n^{2} - 1) + \frac{I}{\pi b R^2} C|n| \nonumber\\
\end{eqnarray}
is the linear growth rate, with electro-osmotic viscosity contrast
\begin{eqnarray}
\label{A}
A= \frac{K_{1}^2-K_{2}^2-(\sigma_{1}+\sigma_{2})(M_{1}-M_{2})}{(K_{1}+K_{2})^2-(\sigma_{1}+\sigma_{2})(M_{1}+M_{2})}.
\end{eqnarray}
Additionally,
\begin{eqnarray}
\label{B}
B= \frac{M_{1}K_{2}^2 + M_{2}K_{1}^2-(\sigma_{1}+\sigma_{2})M_{1}M_{2}}{(K_{1}+K_{2})^2-(\sigma_{1}+\sigma_{2})(M_{1}+M_{2})},
\end{eqnarray}
and
\begin{eqnarray}
\label{C}
C= \frac{M_{1}K_{2} - M_{2}K_{1}}{(K_{1}+K_{2})^2-(\sigma_{1}+\sigma_{2})(M_{1}+M_{2})}.
\end{eqnarray}

The second-order mode-coupling terms appearing on the right-hand side of Eq.~(\ref{result}) are given by
\begin{eqnarray}
\label{F}
F(n, n')&=& \frac{|n|}{R} \Bigg \{ \Bigg . \frac{Q}{2 \pi b R^{2}} \Bigg [ \Bigg . A \left [ \frac{1}{2} - {\rm sgn}(nn') \right ] +(DA - E) |n'|[ 1 - {\rm sgn}(nn') ]  \Bigg ] - \frac{\gamma}{R^{3}} \Bigg [ \Bigg . B \left [ 1 - \frac{n'}{2} ( 3 n' + n )  \right ] \nonumber \\
&+& (f+BD) |n'| ({n'}^{2} - 1)[ 1 - {\rm sgn}(nn') ] \Bigg ] +\frac{IC}{\pi b R^2} \Bigg [D |n'|[ 1 - {\rm sgn}(nn') ] - \frac{1}{2}\Bigg ]  \Bigg \},
\end{eqnarray}
and
\begin{eqnarray}
	\label{G}
	G(n,n')= \frac{1}{R}    \left \{  A|n| [ 1 - {\rm sgn}(nn') ] - 1 \right \} ,
\end{eqnarray}
where
\begin{eqnarray}
	\label{D}
	D=  \frac{K_{1}^2-K_{2}^2-(\sigma_{1}-\sigma_{2})(M_{1}+M_{2})}{(K_{1}+K_{2})^2-(\sigma_{1}+\sigma_{2})(M_{1}+M_{2})},
\end{eqnarray}
\begin{eqnarray}
	\label{E}
	E= \frac{(K_{1}-K_{2})^2-(\sigma_{1}-\sigma_{2})(M_{1}-M_{2})}{(K_{1}+K_{2})^2-(\sigma_{1}+\sigma_{2})(M_{1}+M_{2})},
\end{eqnarray}
\begin{eqnarray}
	\label{f}
	f=\frac{M_{1}K_{2}^2 - M_{2}K_{1}^2+(\sigma_{1}-\sigma_{2})M_{1}M_{2}}{(K_{1}+K_{2})^2-(\sigma_{1}+\sigma_{2})(M_{1}+M_{2})},
\end{eqnarray}
and the sgn function equals $\pm 1$ according to the sign of its argument.

Expressions (\ref{result})-(\ref{f}) are the second-order mode-coupling equations describing the time evolution of the interfacial shape during a radial injection process in a Hele-Shaw cell, taking into account electro-osmotic effects. Note that after neglecting the second-order terms in Eq.~(\ref{result}), one verifies that Eqs.~(\ref{result})-(\ref{f}) agree with the simpler linear (first-order) expression previously derived in Refs.~\cite{Bazant1,Bazant2}. In the absence of electro-osmotic effects ($K_{1}=K_{2}=0$), our second-order results also reproduce the second-order expressions obtained in Ref.~\cite{ST4} which analyzed radial injection in a Hele-Shaw cell. 

In order to plot the weakly nonlinear temporal evolution of the interfaces presented in Fig. (2) [manuscript] and Supplementary Fig. (1), we first consider the nonlinear coupling of all the Fourier modes in the interval $2 \leq n \leq 20$, and rewrite the net interfacial perturbation $\delta(\theta,t)$ in terms of the real-valued cosine $a_{n}(t)=\delta_{n}(t)+\delta_{-n}(t)$, and sine $b_{n}(t)=i[\delta_{n}(t)-\delta_{-n}(t)]$ amplitudes. Once this is done, the shape of the evolving interface can be easily acquired through
\begin{eqnarray}
\label{posinter}
{\cal R}(\theta, t)=R(t)+\delta_{0}+\sum_{n=2}^{20}[a_{n}(t)\cos(n\theta)+b_{n}(t)\sin(n\theta)], \nonumber\\
\end{eqnarray}
where $\delta_{0}$ is an intrinsically nonlinear constraint related to the inner fluid mass conservation, as indicated at the beginning of this section. The time evolution of the mode amplitudes $a_{n}(t)$ and $b_{n}(t)$ can be obtained by numerically solving the corresponding coupled nonlinear differential equations
\begin{eqnarray}
\label{modosacopladosa}
%\begin{aligned}
\dot{a}_n =\lambda(n) a_n + \frac{1}{2} \sum_{n'>0} \Big\{W_{0}(n,-n') a_{n'} a_{n+n'} + W_{0}(n,n') a_{n'} a_{n-n'} + W_{0}(n,-n') b_{n'} b_{n+n'} - W_{0}(n,n') b_{n'} b_{n-n'} \Big \},
\end{eqnarray}
and
\begin{eqnarray}
\label{modosacopladosb}
%\begin{aligned}
\dot{b}_n =\lambda(n) b_n  + \frac{1}{2} \sum_{n'>0} \Big\{W_{0}(n,-n') a_{n'} b_{n+n'} + W_{0}(n,n') a_{n'} b_{n-n'} - W_{0}(n,-n') b_{n'} a_{n+n'} + W_{0}(n,n') b_{n'} a_{n-n'}  \Big \},
%\end{aligned}
\end{eqnarray}
where
\begin{equation} 
W_{0}(n , n')=F(n,n')+\lambda (n') G(n,n').
\end{equation}
Eqs.~(\ref{modosacopladosa}) and~(\ref{modosacopladosb}) are obtained by utilizing Eq.~(\ref{result}).

{\bf Initial conditions:} We set the initial ($t = 0$) amplitude of all perturbation modes as $R_{0}/600$, where $R_{0}=R(t=0)$ is the initial unperturbed radius of the interface. To make the initial conditions as general as possible, we consider the action of random phases attributed to each participating sine and cosine mode. This guarantees that the interfacial behaviors we detect are spontaneously generated by the weakly nonlinear dynamics and not by artificially imposing large initial amplitudes for the modes. We stress that these initial conditions are only imposed for the weakly nonlinear plots, while the boundary integral simulations are generated considering different circumstances, as explained in the next section.

\section*{Supplementary Note 2: Fully nonlinear boundary integral method}
\label{FNL}

In the advanced-time regime of the dynamics, where the viscous fingering instabilities are comparable to the interfacial radius, the time evolution of the fluid-fluid boundary can only be appropriately described by fully nonlinear computational methods such as the boundary integral scheme presented in the following paragraphs. According to potential theory~\cite{book}, the solution of Laplace's equation can be written in terms of boundary integrals. Since the velocity $\phi_j$ and current density $\psi_j$ potentials obey Laplace's equation and have continuous normal derivatives [kinematic boundary conditions - Eqs.~(5) and~(6) of the manuscript] across the interface, these two harmonic functions can be constructed as double layer potentials
\begin{eqnarray}
	\label{DLP1}
	\phi(\mathbf{x})&=&\frac{1}{2\pi}\int \mu_1(\mathbf{y})\frac{\partial\ln |\mathbf{x}-\mathbf{y}|}{\partial \mathbf{n(y)}}ds(\mathbf{y})+\frac{Q}{2\pi b}\ln|\mathbf{x}|,~~
\end{eqnarray}	
and
\begin{eqnarray}
	\label{DLP2}	
	\psi(\mathbf{x})&=&\frac{1}{2\pi}\int \mu_2(\mathbf{y})\frac{\partial\ln |\mathbf{x}-\mathbf{y}|}{\partial \mathbf{n(y)}}ds(\mathbf{y})+\frac{I}{2\pi b}\ln|\mathbf{x}|,~~
\end{eqnarray}
respectively, where $\mu_1(\mathbf{y})$ and $\mu_2(\mathbf{y})$ are dipole densities on the interface, $s$ denotes the interface arclength, and $\mathbf{x}$ represents the position vector with the origin located at the center of the cell. Note that the kinematic boundary conditions are automatically satisfied by these potentials. Now, we rewrite the right-hand sides of Eqs.~(1) and~(2) of the manuscript in terms of the double layer potentials~(\ref{DLP1}) and~(\ref{DLP2}), and substitute the resulting expressions in the boundary conditions [Eqs.~(3) and~(4) of the manuscript] to obtain the integral equations
\begin{eqnarray}
	&&\left(\frac{\sigma_1}{H_1}+\frac{\sigma_2}{H_2}\right)\mu_1(\mathbf{x})+\frac{1}{\pi}\left(\frac{\sigma_1}{H_1}-\frac{\sigma_2}{H_2}\right)\int \mu_1(\mathbf{y})\frac{\partial\ln |\mathbf{x}-\mathbf{y}|}{\partial \mathbf{n(y)}}ds(\mathbf{y})\nonumber\\ 
	&-&\left(\frac{K_1}{H_1}+\frac{K_2}{H_2}\right)\mu_2(\mathbf{x})-\frac{1}{\pi}\left(\frac{K_1}{H_1}-\frac{K_2}{H_2}\right)\int \mu_2(\mathbf{y})\frac{\partial\ln |\mathbf{x}-\mathbf{y}|}{\partial \mathbf{n(y)}}ds(\mathbf{y})\nonumber\\
	&=&2\gamma\kappa-\left(\frac{\sigma_1}{H_1}-\frac{\sigma_2}{H_2}\right)\frac{Q}{2\pi b}\ln|\mathbf{x}|^2+\left(\frac{K_1}{H_1}-\frac{K_2}{H_2}\right)\frac{I}{2\pi b}\ln|\mathbf{x}|^2, 
	\label{integr1}
\end{eqnarray}
and
\begin{eqnarray}
	&-&\left(\frac{K_1}{H_1}+\frac{K_2}{H_2}\right)\mu_1(\mathbf{x})-\frac{1}{\pi}\left(\frac{K_1}{H_1}-\frac{K_2}{H_2}\right)\int \mu_1(\mathbf{y})\frac{\partial\ln |\mathbf{x}-\mathbf{y}|}{\partial \mathbf{n(y)}}ds(\mathbf{y}) \nonumber\\
	&+&\left(\frac{M_1}{H_1}+\frac{M_2}{H_2}\right)\mu_2(\mathbf{x})+\frac{1}{\pi}\left(\frac{M_1}{H_1}-\frac{M_2}{H_2}\right)\int \mu_2(\mathbf{y})\frac{\partial\ln |\mathbf{x}-\mathbf{y}|}{\partial \mathbf{n(y)}}ds(\mathbf{y})\nonumber\\
	&=&\left(\frac{K_1}{H_1}-\frac{K_2}{H_2}\right)\frac{Q}{2\pi b}\ln|\mathbf{x}|^2-\left(\frac{M_1}{H_1}-\frac{M_2}{H_2}\right)\frac{I}{2\pi b}\ln|\mathbf{x}|^2,
	\label{integr2}
\end{eqnarray}
where $\displaystyle H_j=\sigma_jM_j-K_j^2$ (with $j=1, 2$). These equations form a system of well-defined Fredholms integral equations of the second kind, and we solve it for the dipole densities $\mu_1(\mathbf{x})$ and $\mu_2(\mathbf{x})$ via the iterative method GMRES~\cite{GMRES}. Once $\mu_1(\mathbf{x})$ and $\mu_2(\mathbf{x})$ are solved, we compute the normal velocity of the interface via Dirichlet-Neumann mapping~\cite{LapMCD}  
\begin{equation}
	V(t)=\frac{1}{2\pi}\int_{}\mu_{1s'}\frac{(\mathbf{x}-\mathbf{x}')^\perp\cdot\mathbf{n(x)}}{|\mathbf{x}-\mathbf{x}'|^2}ds'(\mathbf{x'})+ \frac{Q}{2 \pi b}\frac{\mathbf{x}\cdot\mathbf{n}}{|\mathbf{x}|^2},\label{V1}
\end{equation}
where the subscript $s'$ denotes the partial derivative with respect to arclength and ${\textbf{x}}^{\perp}=(x_2,-x_1)$. Finally, the interface is evolved through
\begin{equation}
	\frac{d{\mathbf{x}}}{d t}\cdot \mathbf{n}=V.
\end{equation}

Note that in Eq.~(\ref{V1}), the normal velocity decreases as the interface size $|\bf x|$ gets large. It prohibits one from computing the dynamics of an interface at very long times. Thus, we apply the rescaling idea~\cite{ShuwangJCP, Zhao17, Zhao18,MengJFM} to accelerate this slow dynamics. We first introduce 
\begin{equation}
	\mathbf{x}=\bar R(\bar t)\mathbf{\bar x}(\bar t,\theta),
	\label{xs}
\end{equation}
and
\begin{equation}
	\bar t=\int_{0}^t \frac{1}{\rho(t^\prime)}dt^\prime,
	\label{ts}
\end{equation} 
where the space scaling  $\bar R(\bar t)$ represents the size of the interface, $\bar{\textbf{x}}$ is the position vector in the scaled interface, and $\theta$ parameterizes the interface. The time scaling function $\rho(t)=\bar{\rho}(\bar{t})$ maps the original time $t$ to the new time $\bar t$. In addition, note that $\rho(t)$ has to be positive and continuous. The evolution of the interface in the scaled frame can be accelerated \cite{ShuwangJCP,Zhao17} or decelerated \cite{Zhao18,MengJFM} by choosing differents $\rho(t)$. A straightforward calculation shows that the normal velocity in the new frame is expressed as
\begin{equation}
	\bar{V}(\bar t)=\frac{\bar{\rho}}{\bar R}V(t(\bar t))-\frac{\mathbf{\bar x}\cdot \mathbf{n}}{\bar R}\frac{d\bar R}{d\bar t}.
	\label{Vrelation}
\end{equation}

In the rescaled frame, we require that the area enclosed by the interface remains constant $\bar{A}(\bar t)=\bar{A}(0)$. Thus, the integration of the normal velocity along the interface in the scaled frame vanishes, i.e., $\displaystyle \int_{\bar{\Gamma}(\bar t)}\bar{V}d\bar{s}=0$.
As a consequence, 
\begin{equation}
	\frac{d\bar R}{d\bar t}=\frac{\bar{\rho}\bar{Q}}{2b\bar{A}(0)\bar R}.
\end{equation}
By choosing $\displaystyle \rho(\bar{t})= \bar R^2(\bar t)$, the interface grows exponentially in the rescaled frame as
\begin{equation}
	\bar{R}(\bar{t})=\exp\left[\frac{\bar{Q}}{2b\bar{A}(0)}\bar t\right]. 
\end{equation}
Taking $\displaystyle \bar \mu_1(\mathbf{\bar x})=\mu_1(\mathbf{x})\bar R(\bar t)$ and $\displaystyle \bar \mu_2(\mathbf{\bar x})=\mu_2(\mathbf{x})\bar R(\bar t)$, we rewrite the integral equations~(\ref{integr1}) and~(\ref{integr2}) in the rescaled frame as
\begin{eqnarray}\label{integr12}
	&&\left(\frac{\sigma_1}{H_1}+\frac{\sigma_2}{H_2}\right)\bar\mu_1+\frac{1}{\pi}\left(\frac{\sigma_1}{H_1}-\frac{\sigma_2}{H_2}\right)\int \bar\mu_1(\bar{\mathbf{y}})\frac{\partial\ln |\bar{\mathbf{x}}-\bar{\mathbf{y}}|}{\partial \mathbf{n(\bar y)}}d\bar s(\bar{\mathbf{y}})\nonumber\\
	&-&\left(\frac{K_1}{H_1}+\frac{K_2}{H_2}\right)\bar\mu_2-\frac{1}{\pi}\left(\frac{K_1}{H_1}-\frac{K_2}{H_2}\right)\int \bar\mu_2(\bar{\mathbf{y}})\frac{\partial\ln |\bar{\mathbf{x}}-\bar{\mathbf{y}}|}{\partial \mathbf{n(\bar y)}}d\bar s(\bar{\mathbf{y}})\nonumber\\
	&=&2\gamma\bar\kappa-\left(\frac{\sigma_1}{H_1}-\frac{\sigma_2}{H_2}\right)\frac{Q\bar R}{2\pi b}(2\ln \bar R+\ln|\bar{\mathbf{x}}|^2)+\left(\frac{K_1}{H_1}-\frac{K_2}{H_2}\right)\frac{I\bar R}{2\pi b}(2\ln \bar R+\ln|\bar{\mathbf{x}}|^2),
\end{eqnarray}
and
\begin{eqnarray}\label{integr22}
	&-&\left(\frac{K_1}{H_1}+\frac{K_2}{H_2}\right)\bar \mu_1-\frac{1}{\pi}\left(\frac{K_1}{H_1}-\frac{K_2}{H_2}\right)\int \bar \mu_1(\bar{\mathbf{y}})\frac{\partial\ln |\bar{\mathbf{x}}-\bar{\mathbf{y}}|}{\partial \mathbf{n(\bar y)}}d\bar s(\bar{\mathbf{y}})\nonumber\\
	&+&\left(\frac{M_1}{H_1}+\frac{M_2}{H_2}\right)\bar\mu_2+\frac{1}{\pi}\left(\frac{M_1}{H_1}-\frac{M_2}{H_2}\right)\int \bar\mu_2(\bar{\mathbf{y}})\frac{\partial\ln |\bar{\mathbf{x}}-\bar{\mathbf{y}}|}{\partial \mathbf{n(\bar y)}}d\bar s(\bar{\mathbf{y}})\nonumber\\
	&=&\left(\frac{K_1}{H_1}-\frac{K_2}{H_2}\right)\frac{Q\bar R}{2\pi b}(2\ln \bar R+\ln|\bar{\mathbf{x}}|^2)-\left(\frac{M_1}{H_1}-\frac{M_2}{H_2}\right)\frac{I\bar R}{2\pi b}(2\ln \bar R+\ln|\bar{\mathbf{x}}|^2).
\end{eqnarray}
Using Eq. (\ref{V1}), we are able to compute the normal velocity in the rescaled frame
\begin{equation}
	\bar V(\mathbf{\bar x})=\frac{1}{2\pi\bar R}\int\bar \mu_{1\bar{s}}\frac{(\mathbf{\bar x}'-\mathbf{\bar x})^{\perp}\cdot\mathbf{\bar{n}}(\bar{\mathbf{x}})}{|\mathbf{\bar x}'-\mathbf{\bar x}|^2}d\bar{s}'+\frac{\bar{Q}}{2\pi b}\frac{\mathbf{\bar x}\cdot \mathbf{\bar{n}}}{|\mathbf{\bar x}|^2}-\frac{\bar{Q}}{2b\bar{A}(0)}\mathbf{\bar x}\cdot \mathbf{\bar{n}},
	\label{Vb}
\end{equation}
where $\mathbf{\bar x}^{\perp}=(\bar{x}_2,-\bar x_1)$. Then we evolve the interface in the scaled frame through
\begin{equation}
	\frac{d\bar{\textbf{x}}(\bar t,\theta)}{d\bar t}\cdot \textbf{n}=\bar V(\bar t,\theta).
\end{equation}

{\bf Initial conditions:} Here we have modified the initial conditions to 
\begin{equation}
{\cal R}(\theta, 0)=R_{0} + R_{0}/100 \left[ \sin(2\theta) + \cos(3\theta) + \sin(5\theta) + \cos(7\theta) \right].
\end{equation}
This is done because the general initial conditions used to obtain the weakly nonlinear plots are considerably convoluted for the boundary integral method, being very expensive computationally. Although less general than the initial conditions considered in the weakly nonlinear evolution, the initial conditions employed in the numerical simulations are still general enough to investigate dynamical behaviors of the interface during the fully nonlinear regime of the dynamics, as it is composed of a mode mixture of sines and cosines, with all the modes having the same initial amplitude ($R_{0}/100$).

\section*{Supplementary Note 3: Nonlinear dynamics of the non-expanding interfaces}

In the manuscript, Fig. 4 highlighted the morphological transition that happens as one reduces the injection rate. Therefore, here we analyze the pattern-forming dynamics in the complete absence of injection, i.e., $Q(t)=0$, and the non-expanding interface of a viscous oil droplet of radius $R(t)=R_{0}$, surrounded by a mixture of water and glycerol, gets disturbed solely due to the action of an external electric field. In these circumstances, our controlling electric current [Eq.~(11) of the manuscript] reduces to the constant (in time) expression
\begin{eqnarray}
\label{Iconst}
I=\frac{\pi b \gamma B}{CR_{0}}(3{n_{\rm max}^2}-1).
\end{eqnarray}

In order to verify its efficiency during the onset of nonlinearities, in Supplementary Fig.~\ref{WNL1} we plot the weakly nonlinear evolution of the interface using expression~(\ref{Iconst}) for three values of $n_{\rm max}$: 3 [S. Fig.~\ref{WNL1}(a)], 5 [S. Fig.~\ref{WNL1}(b)], and 7 [S. Fig.~\ref{WNL1}(c)], which correspond to the electric currents (a) $I = -12.0$ m A, (b) $I = -34.3$ m A, and (c) $I = -67.6$ m A, respectively. In the top panels of Supplementary Fig.~\ref{WNL1} we depict snapshots of the evolving interfaces for $0\leq t \leq t_{f}$, with the final times (a) $t_f=340$ s, (b) $t_f=66$ s, and (c) $t_f=22$ s, and in the bottom panels we show the corresponding time evolution of the rescaled perturbation amplitudes $|\delta_{n} (t)|/R(t)$ ($2 \leq n \leq 20$), where $\delta_{n} (t) = \sqrt{a_n^2(t)+b_n^2(t)}/2$.

By inspecting the top panels of Supplementary Fig.~\ref{WNL1}, it is very evident that in all three cases, the interface evolves from an initial nearly circular boundary to the desired targeted fingered structure. Note that the fingers formed in these $n$-fold patterns are smooth, and there are no clear signs of nonlinear finger ramification. The visual conclusions provided by the top panels are supplemented by the time evolution of the rescaled perturbation amplitudes depicted in the bottom panels. By scrutinizing these bottom panels, one can observe that during the whole time evolution, the dynamics are dominated by modes (a) 3, (b) 5, and (c) 7, and thus quantitatively supporting the formation of the patterned structures shown in the top panels.

In Supplementary Fig.~\ref{FNL1} we present the fully nonlinear evolution of the interfaces employing our boundary integral method. The patterns are generated considering physical parameters identical to those utilized in Supplementary Fig.~\ref{WNL1}, except for the values of the final times, which are taken as (a) $t_f= 280$ s, (b) $t_f= 68$ s, and (c) $t_f=26$ s.

Analyzing the patterns depicted in Supplementary Fig.~\ref{FNL1} one immediately verifies that the efficiency of the controlling current~(\ref{Iconst}) obtained by means of a linear theory goes beyond the linear and weakly nonlinear dynamics, remaining very useful to set the number of fingers appearing at the interface even at the fully nonlinear regime. Despite the intricate, fully nonlinear morphologies, the fingers formed in the patterns of Supplementary Fig.~\ref{FNL1} show no tendency to split. In addition, we stress that neither of these patterns evolves to a stationary state. On the opposite, the fully nonlinear evolution reveals that the fingers keep growing until their bases meet near the origin, characterizing the onset of interface pinch-off phenomena. This behavior is easily observed in the 5-fold and 7-fold patterns shown in Supplementary Figs.~\ref{FNL1}(b) and~\ref{FNL1}(c), respectively. Note the resemblance between the 7-fold pattern depicted here with the one illustrated in Fig. 4(f) of the manuscript, which highlights that when $Q(t)$ is small, the morphologies are dominated by pinch-off formation.

\renewcommand{\figurename}{Supplementary Fig.}

\begin{figure*}[ht]
	\centering
	\includegraphics[scale=0.62]{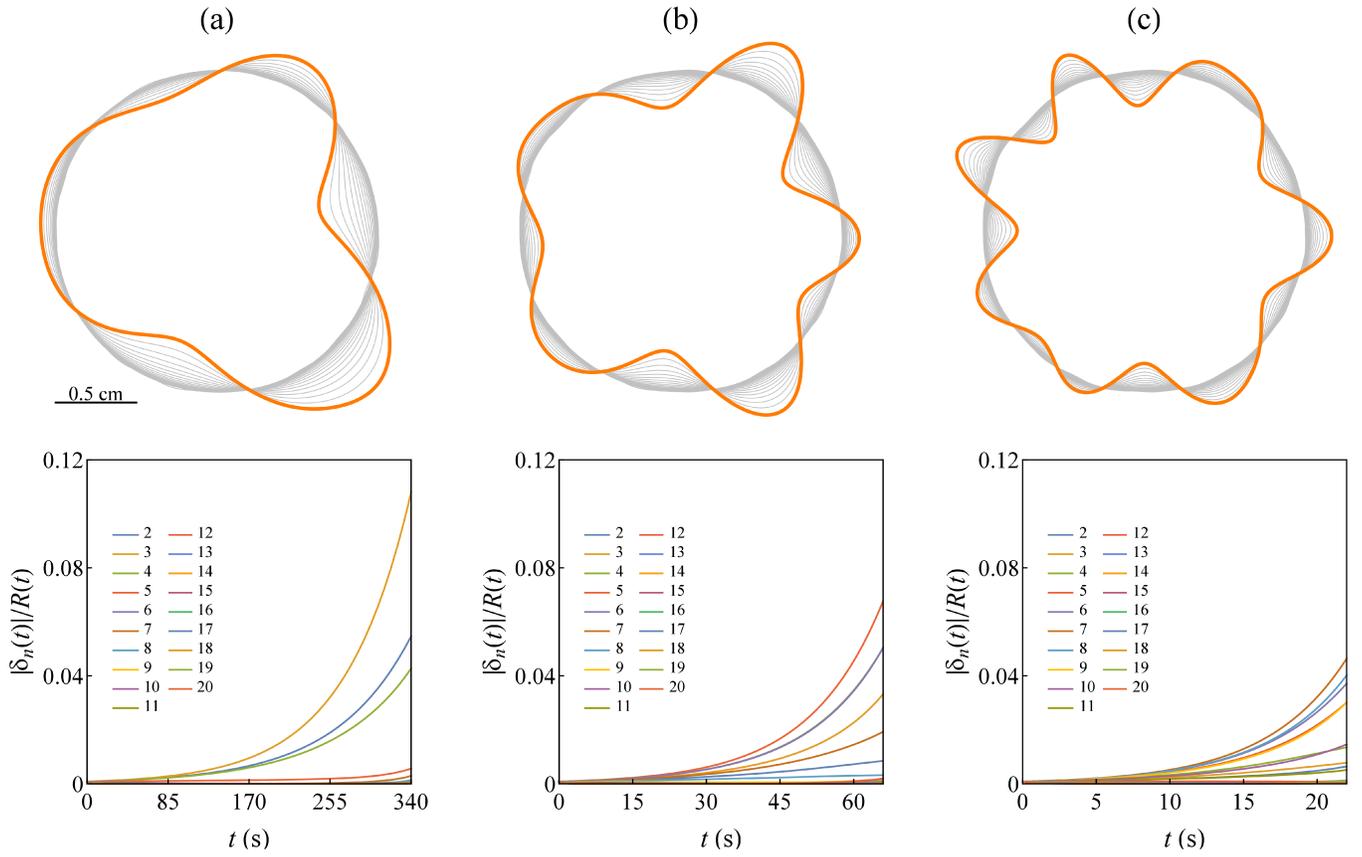} 
	
	\caption{{\bf Perturbative dynamics of non-expanding interfaces.} Weakly nonlinear time evolution of the non-expanding [$Q(t)=0$] interfacial patterns for $0\leq t \leq t_{f}$ generated by utilizing the controlling electric current $I$ [Eq.~(\ref{Iconst})] for (a) $n_{\rm max}=3$ and $t_f=340$ s, (b) $n_{\rm max}=5$ and  $t_f=66$ s, and (c) $n_{\rm max}=7$ and $t_f=22$ s, including modes $2 \leq n \leq 20$ and a random initial phase. The corresponding time evolution of the rescaled perturbation mode amplitudes $|\delta_{n} (t)|/R(t)$ are shown in the bottom panels. The physical parameters are given in the manuscript in Section Methods.}
	\label{WNL1}
\end{figure*}  

\begin{figure*}[ht]
	\centering
	\includegraphics[scale=0.62]{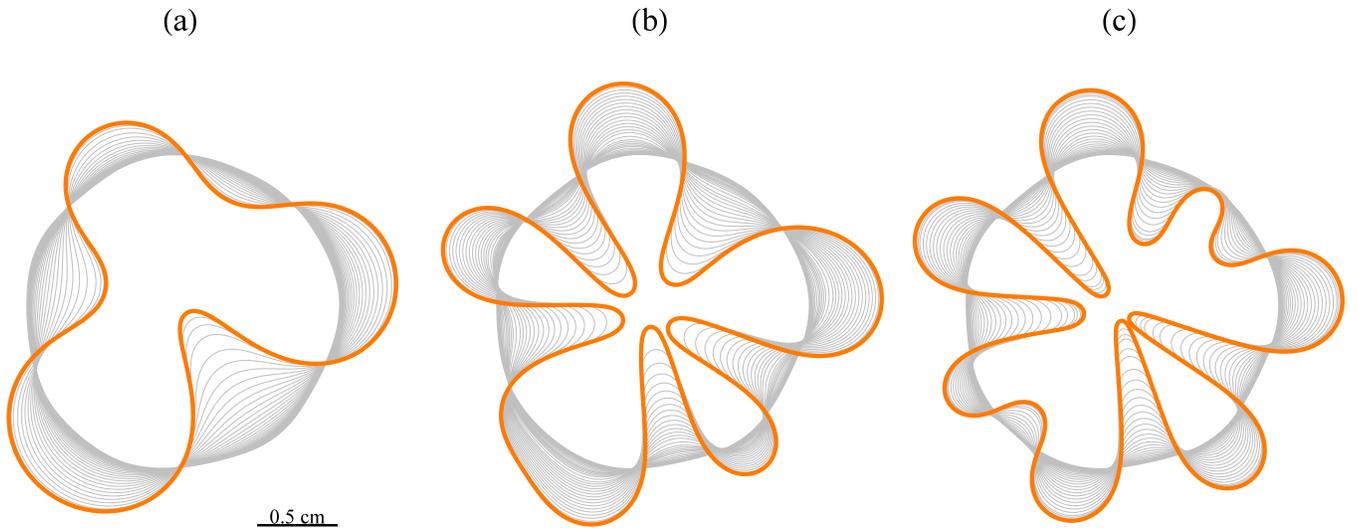} 
	
	\caption{{\bf Numerical simulations of non-expanding interfaces.} Fully nonlinear time evolution of the non-expanding [$Q(t)=0$] interfacial patterns for $0\leq t \leq t_{f}$ generated by utilizing the controlling electric current $I$ [Eq.~(\ref{Iconst})] for (a) $n_{\rm max}=3$ and $t_f= 280$ s, (b) $n_{\rm max}=5$ and  $t_f= 68$ s, and (c) $n_{\rm max}=7$ and $t_f=26$ s.}
	\label{FNL1}
\end{figure*}

%\blankpage
\clearpage

\section*{Supplementary Note 4: Flows performed under constant $I$ and $Q$}

\begin{figure*}[ht]
	\centering
	\includegraphics[scale=0.62]{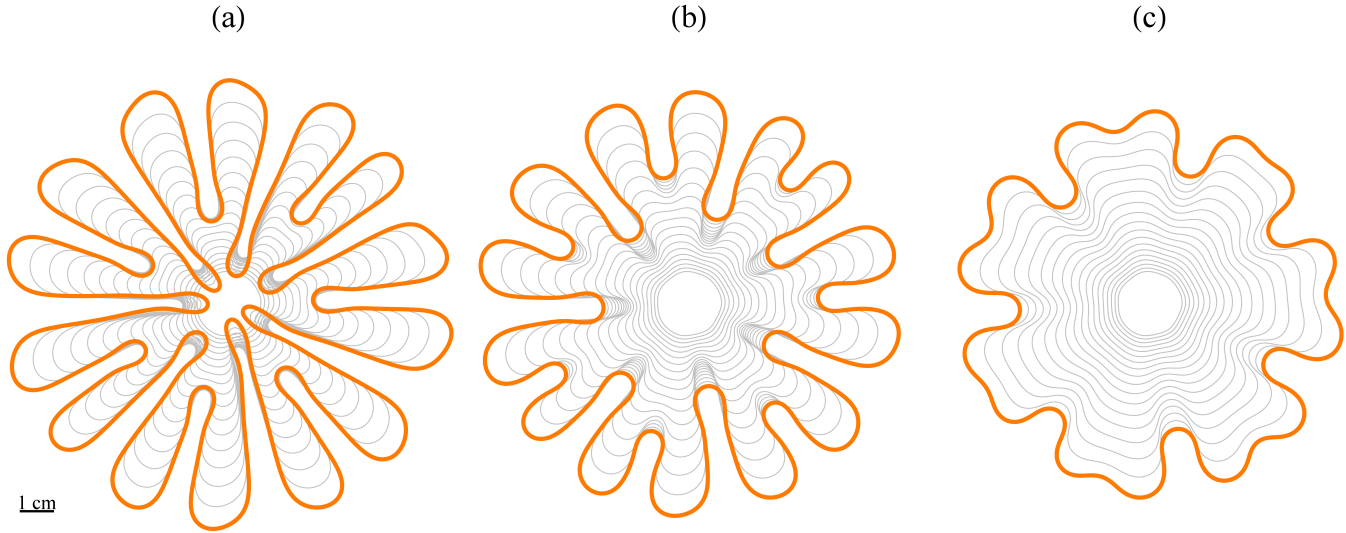} 
	
	\caption{{\bf Uncontrolled, disordered growth of the interface.} Time evolution of the fully nonlinear fluid-fluid interface, illustrating typical fingering patterns during injection-driven, electro-osmotic Hele-Shaw flows, performed under constant injection rate $Q=6.28 \times 10^{-9}$ m$^3$/s and three different values of electric current: (a) $I=-40$ mA, (b) $I=0$ mA, and (c) $I=40$ mA. The final time is taken as $t_f = 144.75$ s. Positive (negative) values of electric current stabilize (destabilize) the interface, in agreement with the experimental results reported in Ref.~\cite{Bazant2}. In addition, note that negative currents favor the emergence of pinch-off events. In all the panels, one observes that new fingers are continuously generated as the interface evolves radially due to nonlinear ramifications.}
	\label{QIconst}
\end{figure*}

\clearpage

\section*{Supplementary Note 5: Analysis of the self-similar growth for $n_{\rm max}=3$}

\begin{figure*}[ht]
	\centering
	\includegraphics[scale=0.90]{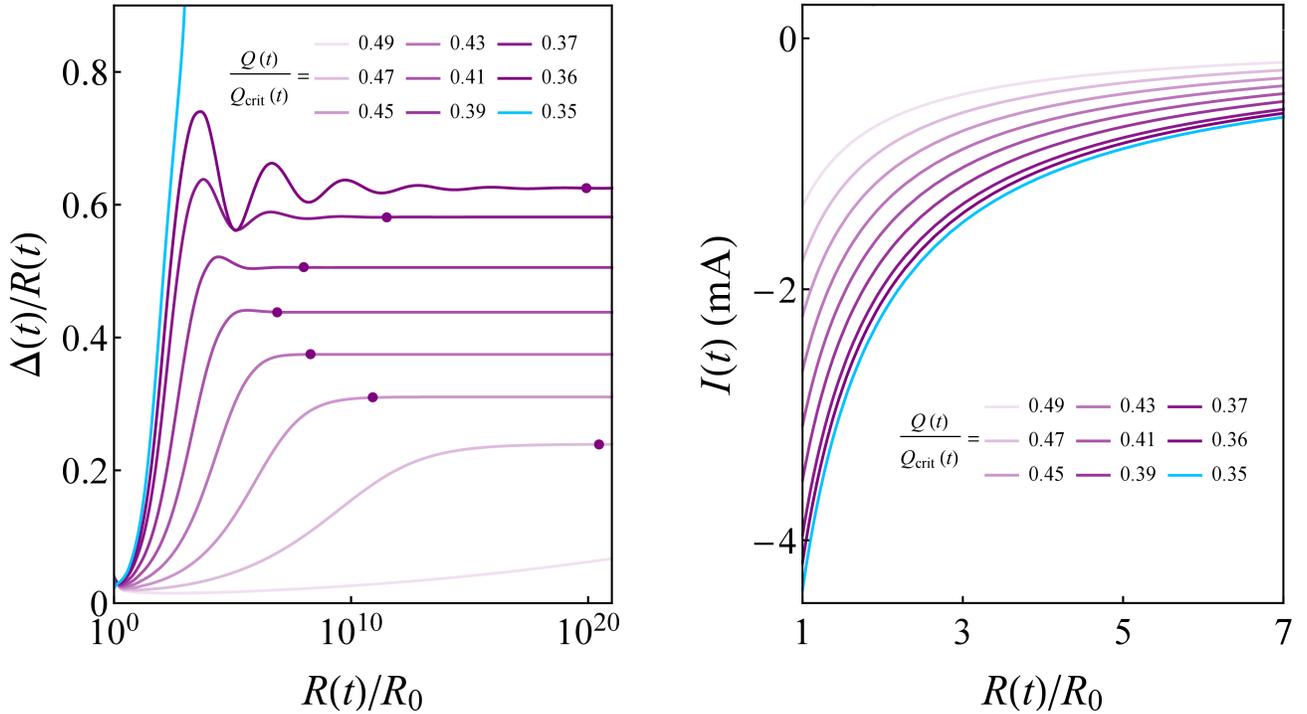} 
	
	\caption{{\bf Analysis of the self-similar growth.} Left panel: Behavior of the shape factor $\Delta(t)/R(t)$ with respect to variations in the ratio $R(t)/R_0$, for flows performed utilizing the controlling electric current [Eq.~(11) of manuscript] with $n_{\rm max}=3$ and various injection rates $Q(t)$. The small dots indicate the points in which the shape factor saturates, i.e., the onset of the self-similar regime. For the curve associated with $Q(t)=0.49 \times Q_{\rm crit}(t)$, the small dot lies outside the plot interval. As for the curve related to $Q(t)=0.35 \times Q_{\rm crit}(t)$, this case do not reach the self-similar evolution. Right panel: Plot of the electric current [Eq.~(11) of manuscript] as a function of $R(t)/R_0$ for the injection rates utilized in the left panel.}
	\label{shapefac}
\end{figure*}

\begin{figure*}[ht]
	\centering
	\includegraphics[scale=0.75]{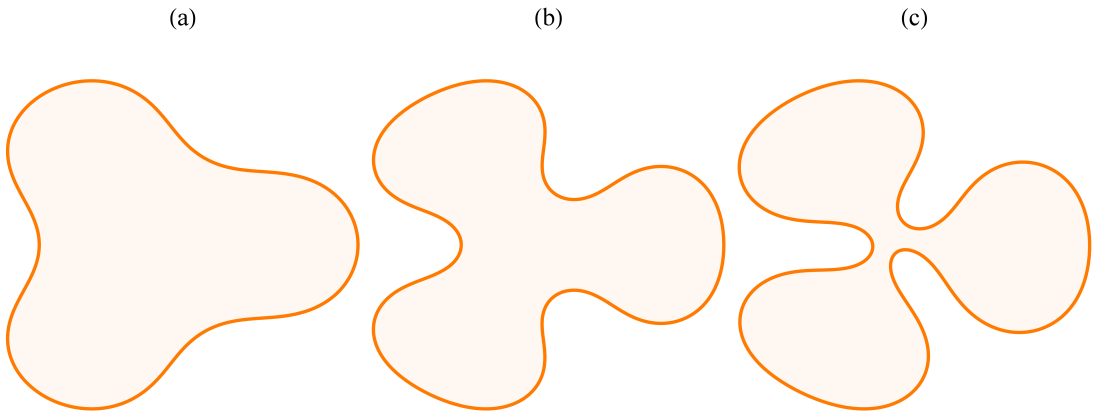} 
	
	\caption{{\bf Snapshots of fully nonlinear interfaces for different injection rates.} Gallery of representative fully nonlinear 3-fold patterns corresponding to the cases (a) $Q(t)=0.45 \times Q_{\rm crit}(t)$, (b) $Q(t)=0.37 \times Q_{\rm crit}(t)$, and (c) $Q(t)=0.35 \times Q_{\rm crit}(t)$, shown in Supplementary Fig.~\ref{shapefac}. The interfaces (a) and (b) are plotted at the onset of the self-similar regime, i.e., for (a) $R/R_{0} = 8.2 \times 10^{10}$ and (b) $R/R_{0} = 3.2 \times 10^{11}$. The pattern depicted in (c) does not reach the self-similar regime, and is plotted for $R/R_{0} = 1.06 \times 10^{3}$.}
	\label{FNL2}
\end{figure*}

\clearpage

\section*{Supplementary Note 6: Selecting other modes $n_{\rm max}$}

\begin{figure*}[ht]
	\centering
	\includegraphics[scale=0.50]{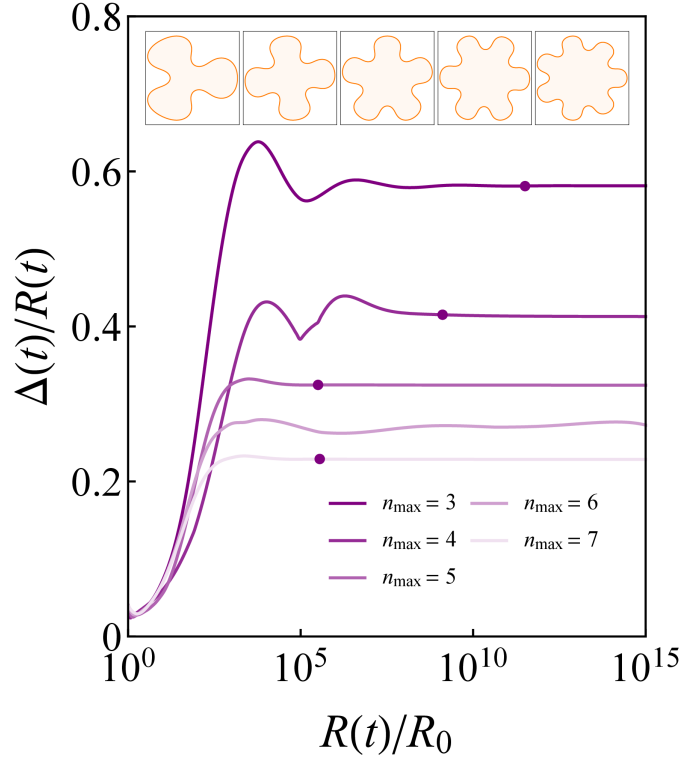} 
	
	\caption{{\bf Selecting the interface symmetry.} Behavior of the shape factor $\Delta(t)/R(t)$ with respect to variations in the ratio $R(t)/R_0$, for flows performed utilizing the controlling electric current [Eq.~(11) of manuscript] with $n_{\rm max}=3, 4, 5, 6,$ and $7$. Here, $Q(t) = 0.37 \times Q_{\rm crit}(t)$, where $Q_{\rm crit}(t)$ is evaluated for each mode $n_{\rm max}$ utilizing Eq.~(13) of manuscript. The small dots indicate the points in which the shape factor saturates, i.e., the onset of the self-similar regime. For the curve associated with $n_{\rm max}=6$, the small dot lies outside the plot interval. The resulting self-similar shapes are depicted as insets.}
	\label{othermodes}
\end{figure*}